\documentstyle[graphicx]{mn2e}

\newif\ifAMStwofonts

\title[The clustering of EROs in UKIDSS DXS Elais-N1]{Clustering of Extremely 
Red Objects in Elais-N1 from the UKIDSS DXS with optical photometry from 
Pan-STARRS 1 and Subaru}
\author[Kim et al.]{Jae-Woo Kim$^{1,2}$\thanks{E-mail:
kjw0704@astro.snu.ac.kr}, Alastair C. Edge$^{1}$, David A. Wake$^{3,11}$, Violeta Gonzalez-Perez
$^{1,4}$, 
\newauthor Carlton M. Baugh$^{1}$, Cedric G. Lacey$^{1}$, Toru Yamada$^{5}$, Yasunori Sato$^{6}$, 
\newauthor William S. Burgett$^{7}$, Kenneth C. Chambers$^{7}$, Paul A. Price$^{8}$, Sebastien Foucaud$^{9,10}$, 
\newauthor Peter Draper$^{1}$ and Nick Kaiser$^{7}$ \\ % and others?
$^{1}$Institute for Computational Cosmology, Department of Physics, University 
of Durham, South Road, Durham DH1 3LE, UK\\
$^{2}$Center for the Exploration of the Origin of the Universe, 
Department of Physics and Astronomy, Seoul National University, Seoul, 
Korea\\
$^{3}$Department of Astronomy, University of Wisconsin, Madison, WI 53706, USA\\
$^{4}$Centre de Physique des Particules de Marseille, Aix-Marseille Universit\'e,  CNRS/IN2P3, Marseille, France\\
$^{5}$Astronomical Institute, Tohoku University, Aoba-Sendai, 980-8578, Japan\\ % new email yamada@astr.tohoku.ac.jp
$^{6}$National Astronomical Observatory of Japan, 2-21-1, Osawa, Mitaka, Tokyo 181-8588, Japan\\
$^{7}$Institute for Astronomy, University of Hawaii, 2680 Woodlawn Drive, Honolulu, HI 96822, USA\\
$^{8}$Department of Astrophysical Sciences, Princeton University, Princeton, NJ 08544, USA\\
$^{9}$Department of Earth Science, National Taiwan Normal University, No 88, Tingzhou Road, Sec. 4, Taipei 11677, Taiwan\\
$^{10}$Institute of Astronomy \& Astrophysics, Academia Sinica, P.O. Box 23-141, Taipei 10617, Taiwan\\
$^{11}$Department of Physical Sciences, The Open University, Milton Keynes, MK7 6AA, UK}
\begin{document}

\date{Accepted ???? ????? ??. Received ???? ????? ??; in original form ???? ????? ??}

\pagerange{\pageref{firstpage}--\pageref{lastpage}} \pubyear{2013}
\maketitle

\label{firstpage}

\begin{abstract}
We measure the angular clustering of 33\,415  extremely red objects (EROs) 
in the Elais-N1 field covering 5.33 deg$^{2}$, which cover the redshift 
range $z=0.8$ to $2$. 
This sample was made by merging the UKIDSS Deep eXtragalactic Survey (DXS) with 
the optical Subaru and Pan-STARRS PS1 datasets.
We confirm the existence of a clear break in the angular 
correlation function at $\sim 0.02^{\circ}$ corresponding to 
$1 h^{-1}$ Mpc at $z\sim1$. We find that redder or brighter EROs are 
more clustered than bluer or fainter ones. 
Halo Occupation Distribution (HOD) model fits 
imply that the average mass of dark matter haloes which 
host EROs is over $10^{13} h^{-1} M_{\odot}$ and that EROs have a bias 
ranging from 2.7 to 3.5. 
Compared to EROs at $z\sim1.1$, at $z\sim1.5$ EROs have a higher bias 
and fewer are expected to be satellite galaxies. 
Furthermore, EROs reside in similar dark matter haloes to those that host 
$10^{11.0} M_{\odot}<M_{*}<10^{11.5} M_{\odot}$ galaxies. 
We compare our new measurement and HOD fits with the 
predictions of the {\tt GALFORM} semi-analytical galaxy formation model. 
Overall, the clustering predicted by {\tt GALFORM} gives an 
encouraging match to our results. However, compared to our deductions 
from the measurements, {\tt GALFORM} puts EROs into lower mass haloes and 
predicts that a larger fraction of EROs are satellite galaxies. 
This suggests that the treatment of gas cooling may need to be revised in 
the model.
Our analysis illustrates the potential of clustering analyses to 
provide observational constraints on theoretical models of galaxy 
formation. 

\end{abstract}


\section{Introduction}

In the Lambda Cold Dark Matter ($\Lambda$CDM) paradigm, small fluctuations 
in the primordial density field grow through gravitational instability and 
merge to form more massive structures. In this scenario, small haloes 
in general form first and become the seeds for larger haloes; galaxies 
form at the centres of these dark matter haloes, as the baryons collapse, cool and form stars 
(White \& Rees 1978). Therefore the formation and 
evolution of galaxies depends critically on the properties of dark matter haloes (e.g., Eke et 
al. 2004; Baugh 2006). In addition, the spatial distribution of galaxies must be 
related to that of the underlying dark matter haloes.

In this context, the measurement of the clustering of galaxies makes it 
possible to link galaxy properties with halo properties, since 
the clustering of galaxies is determined by the clustering of their
host haloes, with more massive haloes showing higher clustering amplitudes
than lower mass haloes (e.g., Mo \& White 1996). The most popular 
methods linking them are the two-point correlation function of galaxies 
(Peebles 1980) and the halo model with the Halo Occupation Distribution 
(HOD) framework (Jing, Mo \& Boerner 1998; Benson et al. 2000; Ma \& Fry 2000; 
Peacock \& Smith 2000; Seljak 2000; Scoccimarro et al. 2001; Berlind \& Weinberg 2002; 
for a review see Cooray \& Sheth 2002). 
The two-point correlation function describes the excess probability over random of the existence 
of a galaxy pair at a specific separation. 
The HOD quantifies the probability that 
a certain type of galaxy is hosted by a halo of a given mass. Given a 
cosmology and the HOD, the galaxy correlation function can be generated.

Recently, wide area surveys have provided an opportunity to measure the clustering 
of different galaxies accurately. From optical imaging and spectroscopic 
surveys, the correlation function of low redshift galaxies selected by their 
luminosity or colour has been measured 
(Norberg et al. 2001, 2002; Zehavi et al. 2002, 2005; Coil et al. 2008; Ross \& 
Brunner 2009; Ross, Percival \& Brunner 2010; Zehavi et al. 2011). In addition, 
the halo properties of luminous red galaxies (LRGs) at $z<1$ were estimated 
using redshift information by Blake, Collister \& Lahav (2008), 
Wake et al. (2008a) and  Sawangwit et al. (2011). The correlation function has also 
been used to study the properties of radio galaxies, quasars and Active Galactic Nuclei (AGN) 
(Wake et al. 2008b; Croom et al. 2005; Coil et al. 2009; Ross et al. 2009; Hickox et al. 2011).

Extracting galaxies at $z>1$, especially red galaxies, is difficult 
since the bulk of their stellar emission is 
redshifted to IR wavelengths. Thus a near-IR dataset is required to 
select red, passive galaxies at $z>1$. There are several colour criteria 
known to be successful at selecting galaxies at $z>1$. Firstly, Extremely Red 
Objects (EROs, Elston, Rieke \& Rieke 1988) can be selected by their red 
optical/near-IR colour (e.g. $(i-K)_{AB}>2.45$). This selection is efficient 
in detecting massive galaxies, $>10^{11}M_{\odot}$, at $z>1$ (Conselice et al. 
2008). Moreover, it is known that EROs are strongly clustered (Daddi et 
al. 2000; Roche et al. 2002, 2003; Brown et al. 2005; Kong et al. 2006, 2009; 
Kim et al. 2011a) and so are expected to 
reside in massive dark matter haloes (Moustakas \& 
Somerville 2002; Gonzalez-Perez et al. 2009; Palamara et al. 2013). However, EROs selected with a simple 
colour cut are contaminated by dusty, star-forming galaxies (Pozzetti \& 
Mannucci 2000; Smail et al. 2002; Roche et al. 2002; Cimatti et al. 2002, 
2003; Moustakas et al. 2004; Sawicki et al. 2005; Simpson et al. 2006; 
Conselice et al. 2008; Kong et al. 2009). Alternatively, a red near-IR colour, 
$(J-K)_{AB}>1.3$, is useful to select Distant Red Galaxies (DRGs) which are 
predominantly intrinsically red galaxies at $z>2$ (Franx et al. 2003). 
Like 
EROs, DRGs are also strongly clustered (Grazian et al. 2006; Foucaud et al. 
2007; Quadri et al. 2008; Guo \& White 2009). 
For both populations, recent results by 
Quadri et al. (2008) for DRGs and Kim et al. (2011a) for EROs showed that their 
angular correlation functions cannot be described by a single power-law. 
This means that the correlation functions of both populations can be  
separated into the contributions from the one-halo term  (the clustering of 
objects in the same halo) and the two-halo term (the clustering of galaxies 
in different haloes). Another intermediate redshift selection successfully 
defined by optical/near-IR 
colours is the $BzK$ selection (Daddi et al. 2004). This population can be 
easily split into star-forming ($sBzK$) and passive ($pBzK$) galaxies at 
$z>1.4$, and are also strongly clustered (Kong et al. 2006; Hartley et al. 2008; 
McCracken et al. 2010; Merson et al. 2013).

Despite the successful colour criteria for selecting high redshift galaxies, 
previous measurements of the correlation function have suffered from the small 
areas surveyed. In particular, the lack of wide-field near-IR imaging data 
has prevented the detection of sufficiently large samples of 
high redshift galaxies. However, this has now been overcome with recent wide and 
deep near-IR surveys using the latest wide field cameras such as the Wide Field Camera 
(WFCAM, Casali et al. 2007). The UKIRT Infrared Deep Sky Survey (UKIDSS, 
Lawrence et al. 2007) is the most comprehensive near-IR survey to date. In 
this paper, we use near-IR images of a wide, contiguous field from the 7th and 
8th Data Release (DR7 \& DR8) of the Deep eXtragalactic Survey (DXS), a  
sub-survey of UKIDSS, in combination with additional optical datasets. From the merged optical to 
near-IR catalogue, the clustering and halo properties of EROs have been 
measured and are discussed.

In \S~2, we describe data analysis methods such as the compilation of 
multi-wavelength datasets, ERO selection method and photometric redshift 
determination. Then the analysis methods used to determine the clustering and 
halo modeling are described in \S~3. We present the results in \S~4, 
and discuss them in comparison with theoretical models in \S~5. 
Unless otherwise noted, the photometry is 
quoted in the AB system. Throughout the bulk of the paper we assume 
the following cosmology : $\Omega_{\rm m}=$0.27, $\Omega_{\Lambda}=$ 0.73, 
$\sigma_{8}=0.8$ and $H_{0}=$ 100 $h$ km s$^{-1}$ Mpc$^{-1}$ with $h=$0.73; the 
exception is in \S~5 in which we adopt a slightly different set of 
cosmological parameters to match those used in a galaxy 
formation simulation which we test against our new measurements. 

\section{Data}

In this section we first describe the near-IR photometry (\S~2.1) and 
supplementary optical photometry (\S~2.2) before discussing the selection 
of EROs (\S~2.3) and the estimation of their photometric 
redshifts (\S~2.4).

\subsection{UKIDSS}
The UKIRT Infrared Deep Sky Survey (UKIDSS) began in 2005 and consists of 5 
sub-surveys (Lawrence et al. 2007). The Wide Field Camera (WFCAM, Casali et 
al. 2007) mounted on the UK Infrared Telescope (UKIRT) has been used to obtain 
UKIDSS images. The Deep eXtragalactic Survey (DXS) is a deep, wide survey 
mapping 35 deg$^2$ with a 5$\sigma$ point-source sensitivity of 
$J_{AB}\sim23.2$ and $K_{AB}\sim22.7$ as one of the sub-surveys. It covers 4 
different fields and aims to detect photometric samples of $z\sim1-2$ galaxies.

WFCAM is composed of four Rockwell Hawaii-II 2K$\times$2K array
detectors (Casali et al. 2007). The sky coverage of each detector is 
13.7$\times$13.7 arcmin$^2$ with 0.4 arcsec/pixel. In order to avoid an 
undersampled point spread function caused by the relatively large pixel scale, 
a microstepping technique is applied, i.e. 0.2 arcsec/pixel for the final 
science image. Since there are gaps between each detector, 4 exposures are 
necessary to generate a contiguous image covering 0.8 deg$^2$.

In this study we deal with the Elais-N1 field centred on $\alpha=$
16$^{h}$ 11$^{m}$ 09.7$^{s}$, $\delta=$ +55$^{d}$ 0$^{m}$ 47.0$^{s}$ (J2000). 
The datasets from UKIDSS data releases 7 and 8 were used for this 
work\footnote{http://surveys.roe.ac.uk/wsa/}. In these releases, 
$K$-band data covers the whole region 
but $J$-band coverage is currently only $\sim$56 per cent. 
The typical seeing is $\sim0.9''$ 
at $J$ and $\sim0.8''$ at $K$. Although the $K$-band dataset has mapped 6.5 deg$^2$ 
after masking unreliable regions, the actual area for this work depends on the
optical datasets (see next subsection). 

There are some known issues about the WFCAM images and database catalogues,
 such as cross-talk and non-optimal galaxy photometry (Dye et al. 2006). To 
avoid these, we therefore created 
our own photometric catalogues from the stacked images. Full details are 
described in Kim et al. (2011a) so we simply summarise the main steps in this 
paper. Stacked images from the UKIDSS standard pipeline were 
combined into individual images 
for each pointing using the Swarp software package (Bertin et al. 2002). 
Then astronomical objects were extracted with SExtractor 
(Bertin \& Arnouts 1996) and a 2-arcsec aperture magnitude for colour and 
the AUTO magnitude for total magnitude were also measured. 
Finally spurious objects such as cross-talk images, diffraction spikes 
and duplicated objects in overlapping regions were 
removed. We found 670\,214 objects and determined the
completeness from an artificial star 
test to be $>90$ per cent at the DXS magnitude goals of 
$J_{AB}=23.2$ and $K_{AB}=22.7$.
The Vega-AB offsets for these bands are: $-0.938$ mag. for $J$ and $-1.900$ mag. for $K$. 

\subsection{Other datasets}

In order to identify EROs we require deep optical
imaging. We use two different datasets for this paper, one that best
matches the DXS area (Pan-STARRS) and one that is deeper but over
a smaller area (Subaru). The combination of the two allows us to
determine the clustering properties of EROs as a function of depth
and area.

\subsubsection{Pan-STARRS}

The Panoramic Survey Telescope and Rapid Response System (Pan-STARRS, Kaiser 
\& Pan-STARRS team 2002) is a large optical survey scanning the whole sky visible 
from Hawaii with $grizy$ filters (Tonry et al. 2012). The science objectives are various, from the Solar system 
astronomy to the distant Universe. The Pan-STARRS prototype telescope (PS1) is 
a 3-year science mission performed by the PS1 Science Consortium\footnote{
http://ps1sc.org/}. The 1.8m PS1 telescope feeds a 1.4 gigapixel camera 
covering a 3.2 degree diameter field of view with $grizy$ filters. 

We use the Medium Deep Survey (MDS) data which comprises ten separate fields. 
The data for the Elais-N1 field (MD08) presented here are from observations
between 2009 and 2010. The stacked images were generated by the Pan-STARRS Image 
Processing Pipeline (IPP). 
The stack IDs of PS1 ranges from 129692 to 129259, with the data label ``MD08.refstack.20100713".
The number of stacked images are more than 80, and the shortest mean exposure time 
is $\sim$ 3 hours at $g$-band. The objects were detected by running SExtractor 
(Bertin \& Arnouts 1996). The flux calibration was performed with the IPP 
synthetic photometry database.
The PS1 catalogue is derived from a single Medium Deep Survey
pointing where the photometric uniformity across the field is
very well calibrated. The rotation of the camera minimises
differences in chip sensitivity as each area of sky is observed
by many different chips so our image depth is homogeneous and
well understood.
The magnitude for the 50 per cent detection limit was 
found to be $i_{AB}\sim25.0$ by matching sources with the deeper Subaru 
catalogue (see below for the Subaru data). The PS1 catalogue was merged with the DXS near-IR catalogue 
(DXS/PS1) by finding the closest object within 1-arcsec. In order to calculate the colour of matched objects, a 3-arcsec 
aperture magnitude from PS1 was used, since the typical seeing of PS1 MDS for 
this field is 
$\sim1.2''$ which is worse than the value at $\sim0.8''$ for DXS. The area 
covered by the DXS/PS1 combination is 5.33 deg$^2$. Galactic extinction was 
corrected for using the dust map of Schlegel, Finkbeiner \& Davis (1998).

\subsubsection{Subaru}

Time was obtained on Subaru to provide a deep comparison for the
UKIDSS DXS dataset in the $i$-band with Suprime-Cam (Miyazaki et al. 2002) and 
a joint catalogue was also produced (DXS/Subaru). 
The Suprime-Cam imaging (PI Yamada) covers part of the 
DXS Elais-N1 field. The images were taken in April 2004 and March 2005. 
The data reduction was made by using SDFRED (Yagi et al. 2002; Ouchi 2004) as well 
as local IDL programs. The standard star of P177D ($\alpha=$
15$^{h}$ 59$^{m}$ 13.6$^{s}$, $\delta=$ +47$^{d}$ 36$^{m}$ 41.8$^{s}$) 
was used for the 
photometric calibration. The 5$\sigma$ point-source limit is $i_{AB}=26.2$ (Sato 
et al., in preparation). However, we cut samples at $i_{AB}=25.5$, 
because there are field-to-field variations
below this level. We determined this limit by
splitting the area into 63 sub-areas of 0.4$\times$0.4~deg$^2$
to establish the variation in the number of objects
detected on a scale smaller than the Suprime-Cam field of view.
For objects in the ranges $23<i_{AB}<23.5$ and $25<i_{AB}<25.5$, the field-to-field variation in counts are 6 and 8 per cent
respectively indicating that the counts are consistent over
this 2 magnitude range. However, for objects with $25.5<i_{AB}<26$
the field-to-field variation rises to 23 per cent which
could significantly affect our clustering on these scales so
we limit our analysis to $i_{AB}<25.5$.
For colour calculations a 2-arcsec aperture 
magnitude was used since the seeing of the Subaru data is similar to 
that of the UKIDSS DXS. The area covered by DXS/Subaru is 3.88 deg$^2$ 
located on the central region of DXS/PS1. 
As for DXS/PS1, Galactic extinction was also corrected for using the 
dust map of Schlegel et al. (1998).
The Vega-AB offsets for these bands are: $-0.39$ mag. for $i$.

\subsubsection{SWIRE}

The Elais-N1 field was also mapped by the Spitzer Wide-area InfraRed 
Extragalactic (SWIRE) survey (Lonsdale et al. 2003). SWIRE imaged 49 deg$^2$ 
split over 6 fields at mid-IR wavelengths. In this work, a 1.9-arcsec aperture 
magnitude of IRAC band data from 
DR2 (Surace et al. 2005) was merged with the other datasets by the same 
scheme mentioned above. 
However only $3.6$ and $4.5$ $\mu$m catalogues were used to 
measure photometric redshifts for the DXS/PS1 dataset, since the shallower depth 
in the longer wavelength regime. 
These bands are labeled as $[3.6]$ and $[4.5]$ respectively.
The Vega-AB offsets for these bands are $-2.820$ mag. for 
$[3.6]$ and $-3.290$ mag. for $[4.5]$.

\subsection{ERO selection}

In this study EROs are selected using the $i-K$ colour from DXS/PS1 and 
DXS/Subaru. Firstly, Galactic stars must be removed 
to avoid contamination. In the case of DXS/PS1 various schemes were applied. 
Bright stars ($K_{AB}<16.3$) were removed using the magnitude difference between 
$K$-band aperture and total magnitudes. Then stellar sequences in $(i-K)$ vs. 
$(g-i)$ and $(r-[3.6])$ vs. $(r-i)$ colour-colour diagrams were extracted. 
These criteria are $(i-K)_{AB}<0.76(g-i)_{AB}-0.85$ and 
$(r-[3.6])_{AB}<2.29(r-i)_{AB}-0.66$.
Stars in the DXS/Subaru data were selected by comparing with 
those in the DXS/PS1. The faintest candidate stars in DXS/Subaru were not removed, 
but this does not affect our analysis because very few of these objects meet
the ERO colour cut. We note that the fraction of faint stars ($i_{AB}>25$) selected as EROs 
is less than 1 per cent, based on the UKIDSS Ultra Deep Survey DR3 catalogue (Simpson et al. 2006).

From the catalogues with Galactic stars removed, colour criteria were applied 
to select EROs. For DXS/PS1 $(i-K)_{AB}>2.45, 2.95$, $i_{AB}<25$ and 
$K_{AB}<22.7$ limits were applied to satisfy the classical colour cut ($I-K)_{vega}>4$) for EROs and match the observed magnitudes. 
For the DXS/Subaru dataset, $(i-K)_{AB}>2.55, 3.05$, $i_{AB}<25.5$ and 
$K_{AB}<22.7$ limits were used. 
The magnitude limits applied for each band are set to ensure we detect objects uniformly across 
the whole area (see the previous section).
The difference in the applied colour cuts  
between the two catalogues arises purely from the magnitude difference 
between PS1 $i$-band and Subaru $i$-band, derived by comparing common objects in 
both catalogues. Hereafter we quote $(i-K)_{AB}=2.45$ and $2.95$ for EROs from both 
datasets instead of $2.55$ and $3.05$.
Using these criteria we selected 17\,250 and 23\,916 EROs with $(i-K)_{AB}>2.45$ 
and 5\,039 and 7\,959 with $(i-K)_{AB}>2.95$ from the DXS/PS1 and DXS/Subaru datasets respectively. 

Fig. 1 shows the number counts for all galaxies and for just EROs. The circles indicate the number counts 
of all galaxies, and triangles and squares are for EROs with 
bluer ($(i-K)_{AB}=2.55$ for DXS/Subaru and $2.45$ for DXS/PS1) and redder 
($(i-K)_{AB}=3.05$ for DXS/PS1 and $2.95$ for DXS/PS1) cuts from this work, respectively. The 
number counts of galaxies 
in Lane et al. (2007, dashed line) and Kong et al. (2006, upper dotted line),  
and EROs in Kong et al. (2006, lower dotted lines) are 
also displayed. The counts from this work are consistent with previous 
studies. However the number counts decrease at fainter magnitudes ($K_{AB}>21$) due to the depth of the optical 
datasets. The shallow depth of the optical datasets prevents the detection of the reddest galaxies. This 
effect is more significant for the DXS/PS1 sample than DXS/Subaru because of 
the more restricted depth of the PS1 dataset.

% figure : number counts
\begin{figure}
\includegraphics[width=8cm]{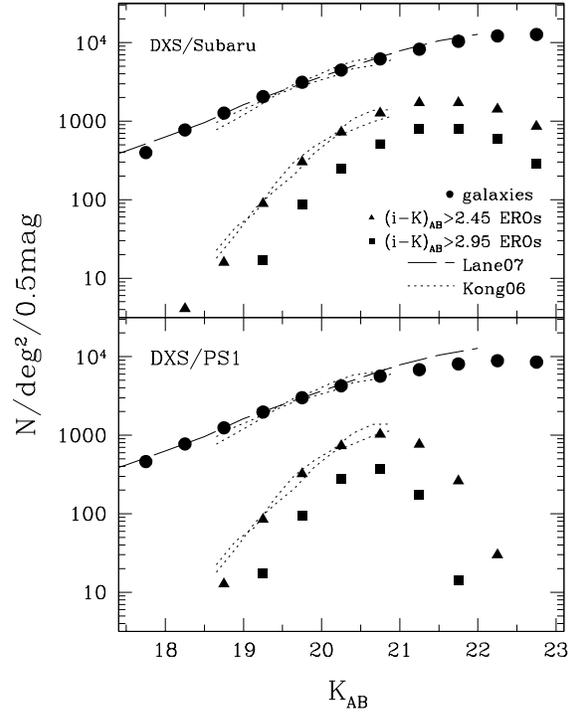}
\caption{Number counts of galaxies and EROs from DXS/Subaru (top panel) and 
DXS/PS1 (bottom panel). The circles are for 
all galaxies in this work, and triangles and squares are for EROs with 
bluer ($(i-K)_{AB}>2.55$ for DXS/Subaru and $2.45$ for DXS/PS1) and 
redder ($(i-K)_{AB}>3.05$ for DXS/Subaru and $2.95$ for DXS/PS1) colours 
respectively. 
The counts in Lane et al. (2007, dashed lines) and Kong et al. (2006, dotted 
lines) are also displayed.  The number counts from the DXS/PS1 sample show 
lower values than those from the DXS/Subaru sample at fainter magnitudes because 
of the more restricted depth of the PS1 dataset. }
\end{figure}

\subsection{Photometric redshift}

The main purpose of this paper is to compare the properties of haloes which 
host EROs at different redshifts by measuring their angular clustering. 
For this purpose the photometric redshifts of EROs were measured 
using the $g$, $r$, $i$, $z$, $J$, $K$, $3.6$ and 
$4.5\mu$m photometric data from DXS-PS1-SWIRE. The EAZY photometric redshift code 
(Brammer, van Dokkum \& Coppi 2008) was run to measure the photometric redshift
of all objects using the default parameters of the EAZY code. To test
the redshift accuracy of this method
we used spectroscopic redshifts from Rowan-Robinson et 
al. (2008), which included those in Berta et al. (2007) and Trichas et al. 
(2010). The normalised median absolute deviation (NMAD) in 
$\Delta z/(1+z_{\rm spec})$ was found to be $\sim0.066$.

However, there are only a small number of spectroscopic redshifts at $z>1$, where most 
EROs are located. Therefore we also applied the empirical method of Quadri \& 
Williams (2010) to constrain the photometric redshift uncertainty for EROs. 
This method assumes that close pairs of galaxies should show a significant 
probability 
of being located at the same redshift. We counted pairs of EROs having an 
angular separations $0.04'<\theta<0.25'$ and those with randomised positions. 
Then, the difference between the two sets in $\Delta z/(1+z_{\rm mean})$ was used to 
measure the photometric redshift uncertainty of EROs. This gave a dispersion 
of $\sigma_{z}\sim0.059$ which is consistent with the NMAD value from spectroscopic 
samples.
Fig. 2 shows the redshift distributions of EROs in DXS/PS1 using a best-fit 
photometric redshift. 
The solid histogram is for $(i-K)_{AB}>2.45$ EROs, and the dashed one is for 
$(i-K)_{AB}>2.95$ EROs. It is apparent that most EROs are located at $z>1$. 
However, the $(i-K)_{AB}>2.45$ ERO selection also contains galaxies at $z<1$.
The median redshifts are 1.176 and 1.291 for $(i-K)_{AB}>2.45$ and $2.95$ EROs 
respectively, and displayed in Fig. 2 as arrows. The trend that redder EROs are 
to be found at higher redshift was also predicted by Gonzalez-Perez et al. (2009).

% figure : redshift distribution of EROs
\begin{figure}
\centering
\includegraphics[width=7cm]{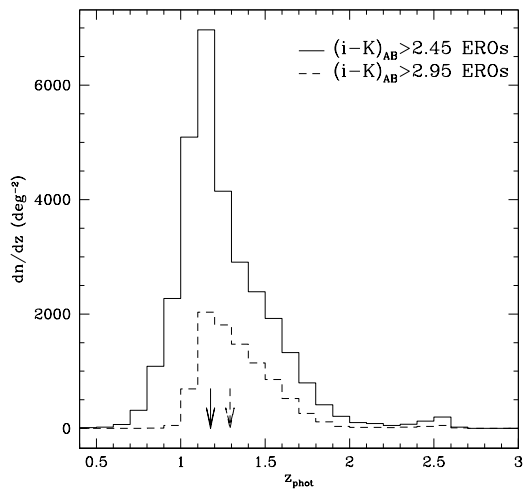}
\caption{The photometric redshift distributions of EROs in DXS/PS1. The 
solid histogram is for $(i-K)_{AB}>2.45$ EROs, and the dashed one is for 
$(i-K)_{AB}>2.95$ EROs. The arrows show the median redshifts of each subsample.}
\end{figure}

\section{Analysis methods}

\subsection{Angular correlation function}

The angular two-point correlation function is the excess probability of finding
a galaxy pair at a given angular separation compared to a random distribution 
(Peebles 1980). We used 
the estimator from Landy \& Szalay (1993) to estimate the angular
two-point correlation function:
\begin{equation}
\omega_{\rm obs}(\theta)=\frac{DD(\theta)-2DR(\theta)+RR(\theta)}{RR(\theta)},
\end{equation}
where DD is the number of observed ERO pairs with separations 
$[\theta,\theta+\Delta\theta]$. For this study we used 
$\Delta \log\theta=$0.15. DR and RR are data-random and random-random pairs 
in the same interval, respectively. 
The random catalogue was generated with 30 times more unclustered 
points than the observed EROs, and had the same angular mask as the EROs. 
All pair counts were normalised to have the same total numbers. 

One of the aims of this study is to investigate the properties of haloes hosting 
EROs at different redshifts. However, our photometric redshift measurement may 
not be accurate enough to split samples cleanly into redshift bins. Therefore 
we used the probability distribution 
function of our photometric redshifts to measure the angular correlation function of 
EROs in different redshift bins and  estimate the redshift distribution function 
of EROs for the halo modeling. The details are described in Wake et al. (2011). 
Briefly, the weight of each ERO is defined as the fractional probability of a 
particular ERO being within a given redshift interval, using the probability distribution 
function from the EAZY code. For the angular correlation function this weight was 
used to count pairs. Also, the weighted probability distribution function of 
photometric redshifts was used to estimate the redshift distribution. This 
strategy is similar to that introduced by Myers et al. (2009). In addition, 
the estimated redshift distribution was also used to measure the number 
density of EROs in each redshift bin using the same scheme as Ross \& Brunner 
(2009).

The error on the correlation function was estimated using the Jackknife 
resampling method to compute the deviation of the correlation functions 
between subfields (for a description see Norberg et al. 2009). We divided the 
whole area into 25 subfields for DXS/Subaru 
and 30 subfields for DXS/PS1, then repeated the measurement of the correlation 
function. From each set of resamplings we can estimate the error using 

\begin{equation}
\sigma^{2}(\theta)=\sum_{i=1}^{N}
\frac{DR_{i}(\theta)}{{DR(\theta)}}
[w_{i}(\theta) - w(\theta)]^{2},
\end{equation}
where $w_{i}$ and DR$_{i}$ are the correlation function and data-random pairs 
excluding the $i^{th}$ subfield, and $N$ is the total number of times that the 
data is resampled. In the Jackknife, $N$ corresponds to the number of 
subfields into which the dataset is divided. 
Then the covariance matrix is calculated with 
\begin{equation}
        \mathbfss{C}_{ij} = (N-1) \langle[w(\theta_{i})
- \overline{w(\theta_{i})}]\cdot[w(\theta_{j})-\overline{w(\theta_{j})}]\rangle, 
\end{equation}
where $\overline{w(\theta_{i})}$ is the mean correlation function of jackknife 
subsamples in the $i$th bin. The covariance matrix was used to fit the halo 
model. 

The restricted survey area leads to a negative offset of the observed 
correlation function from the actual one, which is known as the integral 
constraint (IC, Groth \& Peebles 1977). In order to correct for this bias, we 
applied the same method as Kim et al. (2011a) using the equation in Roche et 
al. (1999),
\begin{equation}
IC=\frac{\sum{RR(\theta)w(\theta)}}{\sum{RR(\theta)}}.
\end{equation}

Since it is known that the correlation function of EROs is not well described by a 
single power-law (Gonzalez-Perez et al. 2011; Kim et al. 2011a), the functional 
form of 
$w({\theta})=\alpha_{1}\theta^{-\beta_{1}}+\alpha_{2}\exp(-\beta_{2}\theta)$ was 
used to describe the correlation function. This form is a close match to the 
angular correlation function measured for ERO samples defined by magnitude and 
colour cut, and is adopted as the 
true underlying correlation function to compute the integral constraint in Eq. 4.
Then Eq. 4 was used to estimate the integral 
constraint, which is then added to our estimate of the angular 
correlation function. Another approach is to use 
the correlation function obtained from the halo model as the 
actual correlation function in Eq. 4 (Wake et al. 2011). For the redshift limited 
samples of EROs, we use the HOD modeled correlation function to calculate the 
integral constraint with the Eq. 4. 
We note that the integral constraint ranges from 0.004 to 
0.008 with the functional form of the correlation function for magnitude or colour 
limited EROs, and that brighter or redder EROs have larger 
integral constraints due to their enhanced clustering strength. 
Similarly the value obtained from the halo model for redshift 
limited samples is between 0.004 and 0.007. We are now probing sufficient area 
that the integral constraint does not have a major effect on our results.

\subsection{Halo modeling}

The halo model (see Cooray \& Sheth 2002 for a review) is widely 
used to estimate the mass of the host dark matter haloes of observed galaxies 
(Blake et al. 2008; Wake et al. 2008a; 
Ross \& Brunner 2009; Zehavi et al. 2011). We apply it to describe the angular 
correlation functions of EROs and hence to measure the properties of the 
dark matter haloes which host EROs.

The halo occupation distribution (HOD) describes the mean number of galaxies 
being hosted by a halo of a given mass ($M$). 
In the halo model, galaxies are separated into centrals and 
satellites. The mean number of galaxies, $N(M)$, is the combination of 
the mean number of central galaxies, $N_{\rm c}(M)$, and satellites, $N_{\rm s}(M)$ 
(Zheng 
et al. 2005; Blake et al. 2008; Wake et al. 2008a; Ross \& Brunner 2009), i.e. 
\begin{equation}
N(M) = N_c(M) + N_s(M),
\end{equation}
where the mean number of the central galaxies and satellites are 
assumed to be described by 
\begin{equation}
N_c(M) = 0.5 \left[ 1 + {\rm erf}\left(\frac{{\log_{10}} (M/M_{\rm cut})}{\sigma_{\rm cut}}\right)\right]
\end{equation}
and 
\begin{equation}
N_s(M) = \left(\frac{M-M_1}{M_0}\right)^{\alpha}, 
\end{equation}
where $M_{\rm cut}$, $\sigma_{\rm cut}$, $M_{0}, M_{1}$ and $\alpha$ are parameters defining 
the shape of HOD. If $M$ is smaller than $M_{1}$, $N_{s}$ is set to 0.

In order to generate the real-space correlation function we followed the 
scheme set out in Ross \& Brunner (2009). Firstly, we modeled the power spectrum 
contributed by galaxies in a single halo (1-halo term) and those in separate 
haloes (2-halo term, $P_{\rm 2 h}$). Also the power spectrum for the 1-halo term is split into 
that of central-satellite pairs ($P_{\rm cs}(k)$) and satellite-satellite pairs 
($P_{\rm ss}(k)$). The equations for each term are 
\begin{equation}
P_{\rm cs}(k) = \int^{\infty}_{M_{\rm vir}(r)} dM n(M) N_c(M)  \frac{2  N_s(M ) u(k|M)}{n^2_{\rm g}},
\end{equation}
\begin{equation}
P_{\rm ss}(k) =\int^{\infty}_{0} dM n(M) N_c(M) \frac{ \left(N_s(M)u(k|M)\right)^2}{n^2_{\rm g}}
\end{equation}
and
\begin{eqnarray}
P_{\rm 2h}(k,r)&=&P_{\rm mat}(k) g^{2} (k, r)  \\
g (k, r)&=&\int_0^{M_{\rm lim}(r)} {\rm d}M n(M) b(M,r) \frac{N(M)}{n_g^{\prime}}u(k|M), \nonumber
\end{eqnarray}
where $n(M)$ is the halo mass function as parameterised in Tinker et al. 
(2010), $u(k|M)$ is the Fourier transform of the halo density profile of 
Navarro, Frenk \& White (1997) and $P_{\rm mat}(k)$ indicates the matter power 
spectrum at the redshift of the sample. 
The term $g(k,r)$ can be thought of as an asymptotic bias. The dependence on 
$r$ arises because only haloes with virial radii less than half of the pair 
separation $r$ of interest are considered; more massive haloes would experience 
an exclusion effect at separation $r$ (see the discussion in Appendix B of 
Tinker et~al. 2005). In addition the virial mass ($M_{vir}(r)$) is defined as 
\begin{eqnarray}
M_{vir}(r) = 200 \times \frac{4}{3}\pi r^{3} \overline{\rho}, \nonumber
\end{eqnarray}
where $\overline{\rho}$ is the mean comoving background density. 
To generate $P_{\rm mat}(k)$ we used the `CAMB' 
software package (Lewis, Challinor 
\& Lasenby 2000) including the fitting formulae of Smith et al. (2003) to 
model non-linear growth.  The average number density of galaxies 
($n_{\rm g}$) is expressed as
\begin{equation}
n_{\rm g} = \int_0^{\infty}{\rm d}Mn(M)N(M).
\end{equation}

\noindent $M_{\rm lim}(r)$, $n_{\rm g}^{\prime}$ and the scale-dependent bias 
($b(M,r)$) are determined using the scheme from Tinker et al. (2005).
The halo bias function ($B(M)$) in Tinker et al. (2010) is used to calculate the 
scale-dependent bias. The calculated power-spectra were converted into 
real-space correlation functions using Fourier transformations

A halo model with three free parameters ($\sigma_{\rm cut}, M_{0}$ and $\alpha$) was 
used to produce the angular correlation function. In this case, $M_{\rm cut}$ was 
fixed by matching the observed number density using the other given parameters, 
and $M_{1}$ was set as $M_{\rm cut}$ which was found to be suitable in previous studies 
(Zehavi et al. 2011). The modeled correlation function was projected to angular 
space using the Limber 
equation (Limber 1954). Then the covariance matrix was used to find the best 
fitting parameters having the minimum $\chi^{2}$ value. The fitting range used was 
$0.001^{\circ}<\theta<0.33^{\circ}$, where the influence of the integral constraint is minimal.

From the fitted parameters, the effective mass ($M_{\rm eff}$), the effective bias 
($b_{\rm g}$) and the satellite fraction can be estimated using 
\begin{equation}
M_{\rm eff} = \int {\rm d}M M n(M)N(M)/n_g,
\end{equation}
\begin{equation}
b_{\rm g} = \int {\rm d}M B(M) n(M)N(M)/n_g
\end{equation}
and
\begin{equation}
f_{\rm sat} = \int {\rm d}M n(M) N_s(M)/n_g.
\end{equation}
In order to determine the properties of haloes hosting EROs at 
different redshifts, we compare all the fitted and estimated values for  
EROs in three redshift bins (see \S~4.3).

\section{Results}
\subsection{Angular correlation function}

% table : ATCF
\begin{table*}
\caption{The amplitudes $A_{\omega}$ and slopes of the correlation functions 
of DXS/Subaru EROs (top 4 rows) for the power-law fit on small 
($0.001^{\circ}<\theta<0.02^{\circ}$) and large ($0.02^{\circ}<\theta<0.33^{\circ}$) scales. 
The bottom 3 rows show same parameters for DXS/PS1 EROs at different redshift 
bins. The number of objects in bottom rows is the sum of weights based on 
photometric redshifts.}
%\label{amplitude}
\centering
\begin{scriptsize}
\begin{tabular}{ccccccc}\\ \hline
Criteria & $A_{\omega}^{small}\times10^{3}$  & $A_{\omega}^{large}\times10^{3}$& slope$^{small}$ & slope$^{large}$ & $\chi^{2}_{small, large}$ & Num.\\
\hline
$(i-K)>2.55, K_{AB}<20.7$ & 1.63$\pm$0.6 &  9.66$\pm$3.5 & 1.10$\pm$0.07 & 0.68$\pm$0.12 & 0.5, 0.3 & 6159\\
$(i-K)>2.55, K_{AB}<21.2$ & 1.42$\pm$0.4 &  7.35$\pm$2.3 & 1.08$\pm$0.05 & 0.69$\pm$0.10 & 1.0, 0.4 & 11726\\
$(i-K)>3.05, K_{AB}<20.7$ & 2.74$\pm$2.2 & 14.45$\pm$6.5 & 1.05$\pm$0.15 & 0.72$\pm$0.15 & 0.6, 0.8 & 2012\\
$(i-K)>3.05, K_{AB}<21.2$ & 3.07$\pm$1.4 & 11.89$\pm$4.2 & 0.98$\pm$0.09 & 0.62$\pm$0.12 & 0.8, 0.3 & 4343\\
                          &              &               &               &               &          &      \\
$1.00<z<1.20, M_{K}<-23$  & 1.93$\pm$0.9 & 10.87$\pm$4.2 & 1.13$\pm$0.09 & 0.70$\pm$0.13 & 0.9, 0.4 & 2272.9\\
$1.15<z<1.45, M_{K}<-23$  & 1.63$\pm$0.5 &  7.81$\pm$2.5 & 1.11$\pm$0.06 & 0.73$\pm$0.10 & 0.9, 0.4 & 3712.5\\
$1.40<z<1.80, M_{K}<-23$  & 3.50$\pm$1.9 & 15.59$\pm$4.6 & 0.91$\pm$0.11 & 0.50$\pm$0.11 & 0.3, 0.4 & 2663.2\\
\hline
\end{tabular}
\end{scriptsize}
\end{table*}

It is known that the clustering properties of EROs depend on magnitude and colour 
(Daddi et al. 2000; Roche et al. 2002, 2003; Brown et al. 2005; 
Georgakakis et al. 2005; Kong et al. 2006, 2009; Gonzalez-Perez et al. 2011; 
Kim et al. 2011a). In this 
section we discuss the properties of the angular two-point correlation 
function of EROs from the DXS/Subaru and the DXS/PS1 samples.
% which are 
%currently the deepest and widest datasets available for ERO studies.

Fig. 3 shows the correlation functions of EROs selected using various criteria. 
All correlation functions in this plot show a clear break at $\sim 0.02^{\circ}$, 
which corresponds to $\sim1 h^{-1}$ Mpc at $z\sim1$ in comoving coordinates.
This break was already reported in Kim et al. (2011a) and implies that a 
single power-law cannot properly describe the correlation function of EROs. 
The presence of significant larger scale clustering was confirmed
by Kim et al. (2011a) with a detailed analysis of multiple sub-areas
of the field and over different ranges in magnitude.
Therefore we tried to fit a power-law ($w(\theta)=A_{w}\theta^{-\delta}$) 
to these correlation functions on small, $0.001^{\circ}<\theta<0.02^{\circ}$, and large, 
$0.02^{\circ}<\theta<0.33^{\circ}$, scales separately. The boundaries for this fitting were 
selected to minimise the influence of the integral constraint on the largest
scales. The values measured are listed in the top four rows of Table 1. 
The amplitudes measured for redder or brighter EROs are larger than found for 
bluer or fainter samples. These features can also be seen in Fig. 3. The top 
two panels display the dependence of clustering on limiting 
magnitude, and the third panel from the top shows the colour dependence, which 
are based on the DXS/Subaru dataset. 

In order to check the consistency of these measurements we compared the 
results with those from the DXS SA22 field
in Kim et al. (2011a) which showed good agreement with previously 
published results. The slopes ($\delta_{small}, \delta_{large}$) in 
Kim et al. were ($0.99\pm0.09, 0.40\pm0.03$) for $K_{AB}<20.7, 
(i-K)_{AB}>2.95$ EROs and ($1.00\pm0.05, 0.51\pm0.02$) for $K_{AB}<20.7, 
(i-K)_{AB}>2.45$ EROs. The values using the same criteria in this work 
are ($1.05\pm0.15, 0.72\pm0.15$) and ($1.10\pm0.07, 0.68\pm0.12$). On 
small scales these values are in agreement within the uncertainty range. 
However, the correlation functions measured in this work are slightly steeper 
than previous results, particularly on the largest scales.
Since Kim et al. used a smaller area, these results might be 
more affected by cosmic variance, explaining the differences on the large 
scales. The most pertinent point is that the correlation function is 
steeper on small scales and flatter on large scales than the single power-law 
with $\delta=0.8$ assumed in most previous studies. Furthermore, to 
compare 
amplitudes directly, we measured the amplitudes again with fixed slopes of 
$\delta=0.99$ and $0.40$ for small and large scales respectively. 
For $K_{AB}<20.7, (I-K)_{AB}>3.05$ EROs, 
the amplitudes ($A_{w}^{small}, A_{w}^{large}$) were 
($4.14\pm0.3, 42.05\pm0.9$)$\times10^{-3}$ in Kim et al., and 
($3.66\pm0.5, 33.26\pm3.7$)$\times10^{-3}$ in this work.  
These are also consistent on small scales but not on the largest scales. 
The reason why the uncertainty 
is larger in this work than the previous one is that we used the Jackknife resampling 
method in this study. It is known that the Poissonian error underestimates the 
uncertainty on the correlation function, especially on the largest scales 
(Ross et al. 2007; Sawangwit et al. 2011; Nikoloudakis, Shanks \& Sawangwit 2013). We also note that 
the Jackknife resampling method works well on large scales (Scranton et al. 2002; Zehavi et al. 2005). 
Based on the scale of inflection and the values of amplitude and slope,
we conclude that our measurements are consistent with previous results
and improve upon them by studying a larger area of sky.

In the bottom panel of Fig. 3, we compare the angular correlation functions 
from the DXS/Subaru and the DXS/PS1 samples. As mentioned above the area for 
DXS/PS1 is larger than that of DXS/Subaru, but DXS/Subaru is deeper. 
The correlation functions of EROs from different optical datasets are well 
matched, suggesting that there is no significant bias for DXS/PS1 
caused by the optical dataset. We use DXS/PS1 EROs to constrain the halo 
properties in the next sections. Overall, the measurements in this work are 
consistent with previous work, although even wider data are necessary to 
measure the angular clustering on large scales more accurately and
to test for field to field variations due to cosmic variance. In the 
near future, PS1 will cover a deeper magnitude range and provide 
more reliable measurements for the clustering of high redshift galaxies
in all four DXS fields.

% figure : ATCF
\begin{figure}
\includegraphics[width=8cm]{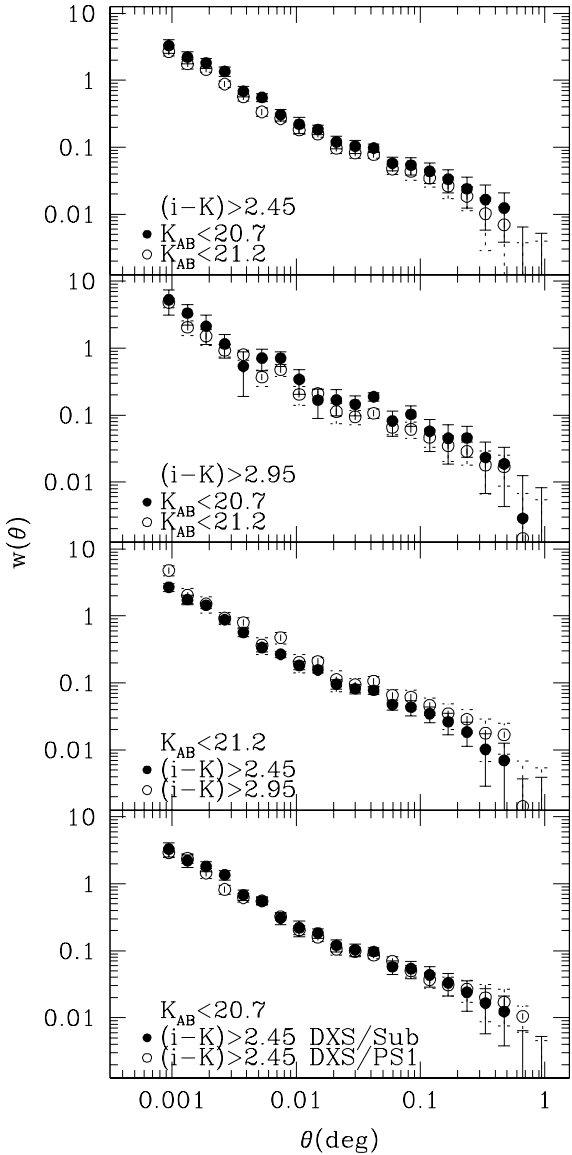}
\caption{Angular two-point correlation functions of EROs selected with various 
criteria (top three panels) based on the DXS/Subaru dataset. The correlation 
functions of EROs from DXS/Subaru and DXS/PS1 are compared in the bottom 
panel. The error on the correlation functions is estimated using the 
Jackknife resampling method.}
\end{figure}

% figure : ATCF
\begin{figure}
\includegraphics[width=8cm]{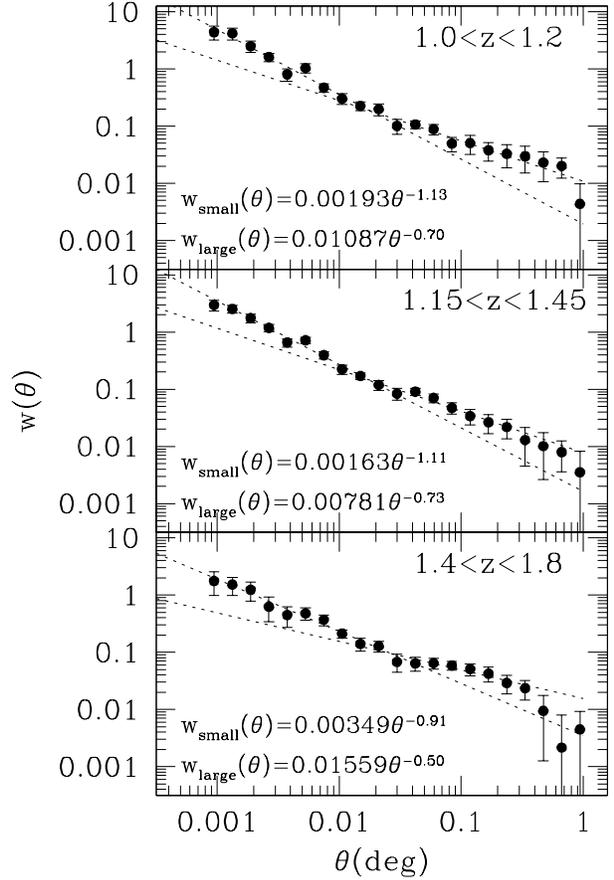}
\caption{Angular two-point correlation functions of DXS/PS1 EROs selected 
by $M_{K}<-23$ and $(i-K)_{AB}>2.45$ at 
$1.0<z<1.2$ (top), $1.15<z<1.45$ (middle) and $1.4<z<1.8$ (bottom). Dotted 
lines indicate the power-law fits on small and large scales separately. 
The equations in each panel are the best power-law fit on small and 
large scales.}
\end{figure}

\subsection{Clustering of EROs in redshift bins}

Photometric redshifts provide an opportunity to compare 
the clustering properties of EROs in different redshift bins. From 
the DXS/PS1 catalogue we classified EROs based on their redshift and 
absolute magnitude. Firstly, we applied a $K$-band absolute magnitude cut 
($M_{K}<-23$) with $(i-K)_{AB}>2.45$ to select EROs with a similar 
$K$-band luminosity in different redshift bins. 
The $K$-band absolute magnitude was calculated using 
the k-correction value obtained from the Bruzual \& Charlot (2003) GALAXEV 
code and a luminosity distance calculated using the Javascript Cosmology 
Calculator (Wright 2006). We assumed a formation redshift of $4<z_{f}<5$ and 
considered a range of metallicities around solar to find a best fit 
to $izJK$ colours of the ERO. The simple stellar population templates with 
a Chabrier (2003) initial mass function were used. Then, the EROs 
were split into three redshift bins : $1.0<z<1.2, 1.15<z<1.45$ and 
$1.4<z<1.8$. The bin size was chosen to select sufficiently large 
samples to enable accurate clustering measurements. 
The angular two-point correlation functions for each sample were measured 
using the probability distribution function of photometric redshift as 
described in \S~3.1. Finally a power-law fit was performed on small and large 
scales separately. The same fitting range as above was used. Fig. 4 
shows the angular two-point correlation functions of DXS/PS1 EROs at 
$1.0<z<1.2$ (top), $1.15<z<1.45$ (middle) and $1.4<z<1.8$ (bottom). Dotted 
lines indicate the power-law fits on small and large scales separately. 
The bottom three rows in Table 1 list the fitted results. The number of 
EROs in each bin is the sum of the weights estimated from the probability distribution 
function of photometric redshift. Although the 
correlation function shows a slightly flatter shape in the highest redshift 
bin than the others, all the estimated power-law slopes have similar values 
within the uncertainty range. 

We also note that the amplitudes on large scales 
with a fixed power-law slope ($\delta=0.7$) are 0.008$\pm$0.0007 at $1.15<z<1.45$ 
and 0.009$\pm$0.0008 at $1.4<z<1.8$. The power-law slope of $0.7$ was derived for 
EROs at $1.0<z<1.2$. These 
similar amplitudes on large scales indicate a higher bias at higher redshift. 
This will be discussed in the next section.

\subsection{Halo modeling}

In this section we study the properties of those haloes hosting EROs at different 
redshifts. The angular correlation functions for DXS/PS1 EROs in different 
redshift bins with $M_{K}<-23$ and $(i-K)_{AB}>2.45$ were used for the halo modeling. The halo 
models were generated at the median redshift of the bins, $z=1.1, 1.3$ and 
$1.5$. The halo model with the three free parameters ($\sigma_{\rm cut}$, $M_{0}$ 
and $\alpha$) mentioned in \S~3.2 was applied at each redshift. We note that 
this assumed 
HOD frame work is basically appropriate for mass or luminosity limited samples. In fact EROs are 
not mass limited samples. Therefore some central or satellite galaxies may be missed, which 
affects the shape of HODs. In this section, we simply apply the standard HOD model to the 
clustering of EROs. Then a modified HOD will be discussed in the next section.

% figure : HOD fit with 3 para
\begin{figure}
\includegraphics[width=8cm]{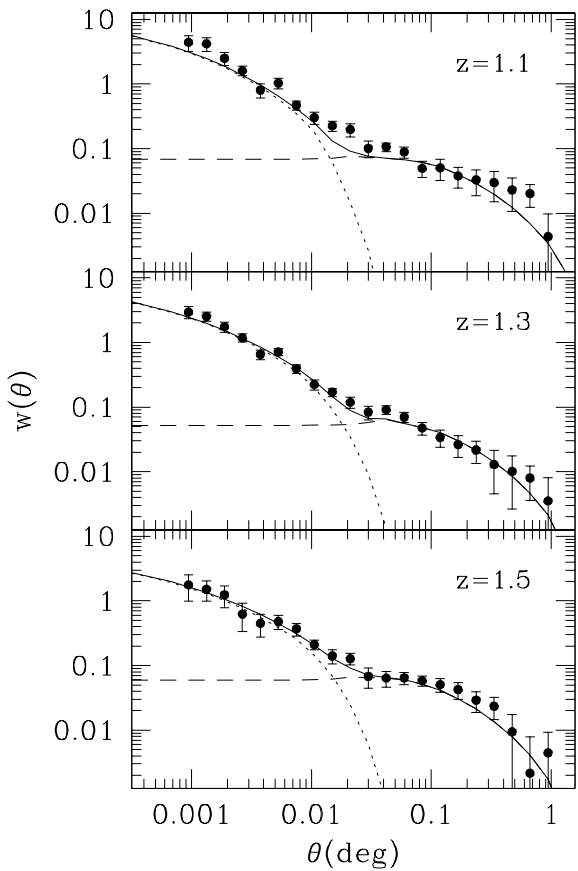}
\caption{Angular correlation functions of $(i-K)_{AB}>2.45$ EROs with $M_{K} < -23$  at different 
redshifts. Circles indicate the correlation function and solid lines are 
best fit halo models. The dotted and dashed lines represent the 1- and 2-halo terms, 
respectively.}
\end{figure}

Fig. 5 shows the angular correlation function estimated from the observed $(i-K)_{AB}>2.45$ EROs, 
brighter than $M_{K}=-23$ in the different redshift bins (circles), and the best fit 
halo models (lines). The dotted and dashed lines represent the 1- and 2-halo terms,
respectively.
The features of the correlation functions of all ERO subsets are relatively well  
fitted by the standard halo model. The HOD fit parameters are 
listed in Table 2. 
The errors on the HOD parameters in the fits were by determined
finding the minimum and maximum
parameter values with $\Delta \chi^2 \le 1$ from the best fit solution.
For other derived parameters ($b_g$, $M_{eff}$ and $f_{sat}$),
$\Delta \chi^2 \le 3.53$ was used.

The top panel in Fig. 6 displays the effective bias which is estimated using 
the equation in \S~3.2. The EROs at higher redshift show a higher bias 
which is similar to the trend found in previous studies of various populations 
(Blake et al. 2008; Wake et al. 2008a; Sawangwit et al. 2011 for Luminous Red 
Galaxies (LRG), Matsuoka et al. 2011; Wake et al. 2011 for stellar mass limited 
samples and Ross, Percival \& Brunner 2010 for absolute magnitude cut samples). 
The biases of EROs reported previously were $\sim3$ at $z=2.1$ for 
$(R-K)_{AB}>3.3$ and $K_{AB}<22.1$ EROs from the 
semi-analytical model in Gonzalez-Perez et al. (2011) and $2.7\pm0.1$ in 
Moustakas \& Somerville (2002), which are similar to our measurements, 
although Moustakas \& Somerville (2002) applied a single power-law and used a 
different criteria, $(I-H)_{AB}>2$ and $H_{AB}<21.9$, for the Las Campanas Infrared 
Survey data (McCarthy et al. 2001; Firth et al. 2002). These values are 
similar to those for low redshift LRGs which have $b_{\rm g}\cong2-3$ (Blake et al. 
2008; Wake et al. 2008a; Sawangwit et al. 2011). If EROs have similar stellar 
masses to LRGs, EROs should be more biased than the lower redshift LRGs. 
However, we note that the 
median stellar mass of an SDSS LRG is $\sim10^{11.5} M_{\odot}$ with a narrow 
distribution (Barber, Meiksin \& Murphy 2007), but in the case of EROs at 
$K_{AB}<21.6$, the distribution shows a peak at $\sim10^{11.3} M_{\odot}$ and 
a sharp cut-off at $\sim10^{11.5} M_{\odot}$ (Conselice et al. 2008). 
Therefore our samples are probably marginally less massive than LRGs at lower redshifts.

The middle panel in Fig. 6 shows the effective mass of dark matter haloes hosting 
EROs. We found that the average halo mass hosting EROs is over 
$10^{13} h^{-1} M_{\odot}$ and that EROs at higher redshift are in slightly 
less massive haloes than those at lower redshift. Gonzalez-Perez et al. (2011) 
reported that 
the median mass of haloes hosting $(R-K)_{AB}>3.3$ and $K_{AB}<20.9$ EROs at 
$z=1.1$ is $4.4\times10^{12} h^{-1}M_{\odot}$ from a semi-analytical model,
and Moustakas \& Somerville 
(2002) estimated the average halo mass hosting EROs as 
$5\times10^{13} h^{-1}M_{\odot}$ at $z\sim1.2$. Recently Palamara et al. 
(2013) published the halo mass of $K_{AB}<18.9$ EROs as $10^{13.09}$ $M_{\odot}$ based on 
the clustering strength with a power law slope of 1.73. Since we have dealt with 
a full halo model in this work rather than a simplified conversion, the result 
may be the better measurement than previous work for EROs.

Other studies use a variety of definitions of ``passive" galaxies at 
$1<z<2$, which 
have differing degrees of overlap with the ERO sample considered in this 
paper. 
Hartley et~al. (2010) used star formation history modelling to 
define galaxies as passive, selecting those in which they inferred 
that the current star formation rate is less than 10\% of the initial 
star formation rate and using a red cut on the rest-frame $U-B$ 
colour. These authors estimated that passive galaxies defined in 
this way with $M_{K}<-23$ at $z<2$ are located in haloes ranging 
from $10^{13} M_{\odot}$ to $5\times10^{13} M_{\odot}$. 
We infer a host halo mass that is slightly higher than the 
prediction of Gonzalez-Perez et~al. (2011), but slightly lower 
than that for the passive galaxies as defined by Hartley et~al. (2010). 
The comparison with the theoretical work is discussed further 
in \S~5. 
However, overall, EROs may reside in slightly less massive haloes 
than passive galaxies of comparable luminosity. It is known 
that EROs can be split on the basis of their colours 
into passive galaxies with old stellar populations or
dusty, star-forming galaxies (Pozzetti \& Mannucci 2000; Smail et al. 2002; 
Roche et al. 2002; Cimatti et al. 2002, 2003; Moustakas et al. 2004; Sawicki 
et al. 2005; Simpson et al. 2006; Conselice et al. 2008; Kong et al. 2009). 
In Kim et al. (2011a), the fraction of old, passive EROs defined in this way 
was found to be more than $\sim60$ per cent. 
In this work it is not possible to split the whole sample into 
these two sub-populations by using $(i-K)$ and $(J-K)$ colours due to the 
lack of $J$-band imaging over the full area. 
So, we simply apply the criterion to the two-colour 
diagram for the region where $J$-band imaging exists. The fraction 
of old, passive EROs is 45.4 per cent at $z=1.1$ and 54.4 per cent at 
$z=1.5$. The significant fraction of dusty, star-forming galaxies may dilute 
the clustering of EROs which might explain why we find lower halo masses 
than pure passive galaxy samples would predict. 

The bottom panel in Fig. 6 shows the evolution of the satellite fraction with 
redshift derived from the HOD model. EROs are made up of a larger fraction of 
satellite galaxies at lower redshifts, although the best fitting 
power-law slope of satellites ($\alpha$) at $z=1.5$ is much larger than unity, 
which is the value typically seen in simulations of galaxy 
formation (e.g. Almeida et~al. 2008) and as recovered in many previous 
analyses of observational measurements of clustering. 
However, in Wake et~al. (2008a), the slope for satellite 
LRGs at $z=0.55$ was $\sim$2.0, and Matsuoka et al. (2011) also reported a similar value for massive galaxies at $0.8<z<1.0$. 
So this is not the first result showing a large slope. 
However, we also note there are other components of the HOD that could lead to a 
low satellite fraction other than a steep satellite slope. The mass thresholds ($M_{\rm cut}$ 
and $M_{0}$) for the HOD may be factors causing the result. The threshold for the HOD of 
central EROs ($M_{\rm cut}$) is smaller than that for LRGs or massive galaxies, which are a few times 
$10^{13} h^{-1} M_{\odot}$. Moreover, $M_{0}$ is larger than SDSS galaxies at $z\sim0.3$ 
(Ross et al. 2010). This means that EROs are in less massive haloes than LRGs when they are the central 
galaxy and in more massive haloes than typical galaxies when they are a satellite. Therefore, these 
effects may lead to the lower satellite fraction than the previous results for ordinary galaxies.

As mentioned above, EROs can be split into old, passive galaxies (OG) or
dusty, star-forming galaxies (DG) on the basis of their $i-K$ and $J-K$ colours. Kim et al. (2011a) 
show that these two sub-populations have very different
clustering properties with OGs being more strongly clustered than DGs.
Unfortunately, the $J$-band coverage of the EN1 field is not complete
so we cannot perform the same analysis as Kim et al. (2011a). Gonzalez-Perez et al. (2011) determine the difference in
clustering of these sub-populations in semi-analytical simulations
and find a comparable difference in the clustering to that
in Kim et al. (2011a). Therefore it is likely that OGs
are more biased and/or in more massive haloes than DGs but
the clustering of DGs is sufficiently similar to that of OGs
that any halo modeling of the combined population is
representative. We will perform a more detailed halo modeling
of each sub-population with the full DXS dataset in a future paper.

Additionally we note the results of halo modeling for stellar mass limited 
samples. Wake et al. (2011) used data from the NEWFIRM medium band survey 
(NMBS; van Dokkum et al. 2009; Brammer et al. 2009; van Dokkum et al. 2010; 
Whitaker et al. 2011) to measure the clustering of stellar mass limited samples of 
galaxies at $1<z<2$. The highest mass limits of their samples were 
$M_{*}=10^{10.7}M_{\odot}$ at $z=1.1$ and $10^{10.78}M_{\odot}$ at $z=1.5$. Comparing our 
results to theirs, the effective mass and bias for EROs are higher than those 
in Wake et al.(2011). This can be easily explained by the higher stellar mass of EROs, as more 
massive or brighter galaxies
reside in more massive haloes (Zehavi et al. 2005, 2011; Foucaud et al. 2010;
Hartley et al. 2010; Matsuoka et al. 2011; Furusawa et al. 2011) and show a
higher bias (Coil et al. 2006; Ross \& Brunner 2009; Zehavi 2011), if we
assume EROs have stellar masses greater than $10^{11} M_{\odot}$. 

Foucaud et al. 
(2010) measured the mass of haloes hosting $10^{11}M_{\odot}<M_{*}<10^{11.5}M_{\odot}$ 
Palomar/DEEP2 galaxies at $1.2<z<1.6$ based on the model in Mo \& White (2002). They 
reported the halo mass and bias as $10^{13.17} h^{-1} M_{\odot}$ and 2.8$\pm$0.6, 
which are similar to our result for EROs at $z=1.3$. Therefore EROs may have similar 
properties with those galaxies.

% figure : measured halo properties
\begin{figure}
\includegraphics[width=8cm]{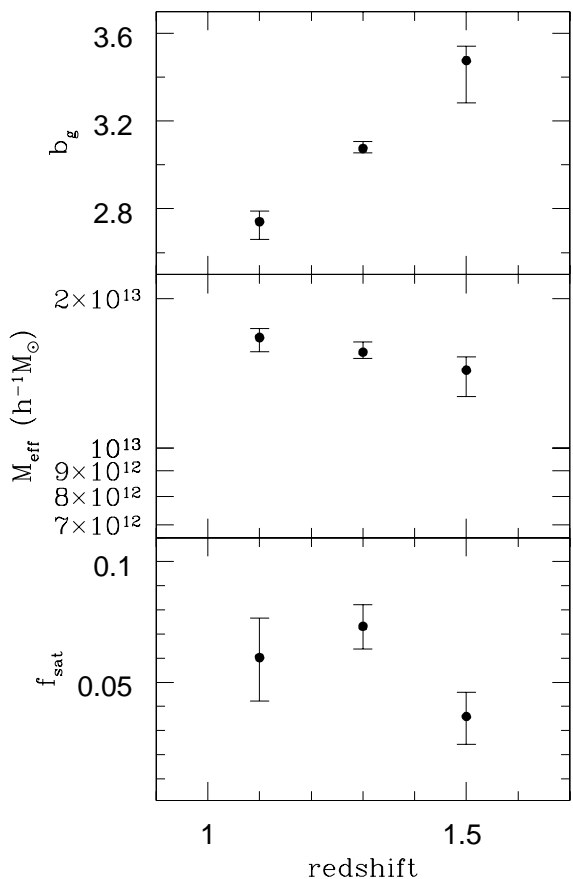}
\caption{The estimated effective bias (top), effective halo mass (middle) and 
satellite fraction (bottom) when the three-parameter halo model is applied to 
the measured angular correlation functions.}
\end{figure}

% table : fitted parameters
\begin{table*}
\caption{The parameters of the best fitting HOD and derived quantities for 
EROs with $M_{K}<-23$ and $(i-K)_{AB}>2.45$ in different 
redshift bins. Column (1) gives the redshift bin, columns (2-5) gives the 
parameters of the best fitting HOD, as defined in Eqs 5-7, columns (6-9) give 
quantities derived from the HOD: column (6) gives the number density of 
galaxies, column (7) lists the effective bias, column (8) the effective 
host halo mass and column (9) gives the fraction of EROs that are inferred 
to be satellite galaxies. The final column describes the quality of the 
HOD fit to the observed angular correlation function, in terms of the 
value of $\chi^{2}$ per degree of freedom.}
\centering
\begin{scriptsize}
\begin{tabular}{cccccccccc}\\ \hline
median & $\sigma_{\rm cut}$  & log$(M_{\rm cut}/h^{-1}M_{\odot})$ & log$(M_{0}/h^{-1}M_{\odot})$ & $\alpha$ & $n_{\rm g}$ & $b_{\rm g}$ & log$(M_{\rm eff}/ h^{-1} M_{\odot})$ & $f_{\rm sat}$ & $\chi^{2}/$dof\\
$z$ & & & & & $(10^{-4}h^{3}{\rm Mpc}^{-3})$ & & & \\

\hline
1.1 & $0.24^{+0.04}_{-0.02}$ &  $12.875^{+0.046}_{-0.025}$ & $14.107^{+0.032}_{-0.022}$ & $1.00^{+0.04}_{-0.03}$ & $ 3.0$ & $2.74^{+0.05}_{-0.08}$ & $13.223^{+0.018}_{-0.029}$ & $0.060^{+0.016}_{-0.018}$ & 2.87\\
1.3 & $0.07^{+0.02}_{-0.03}$ &  $12.745^{+0.009}_{-0.004}$ & $13.701^{+0.010}_{-0.006}$ & $1.64^{+0.10}_{-0.08}$ & $ 3.2$ & $3.07^{+0.03}_{-0.02}$ & $13.193^{+0.021}_{-0.012}$ & $0.073^{+0.009}_{-0.009}$ & 2.71\\
1.5 & $0.19^{+0.08}_{-0.03}$ &  $12.832^{+0.096}_{-0.017}$ & $13.800^{+0.023}_{-0.019}$ & $1.99^{+0.11}_{-0.11}$ & $ 1.9$ & $3.48^{+0.06}_{-0.19}$ & $13.157^{+0.026}_{-0.053}$ & $0.036^{+0.010}_{-0.012}$ & 0.93\\
\hline
\end{tabular}
\end{scriptsize}
\end{table*}

\section{Comparison with galaxy formation models}

EROs are massive galaxies, possibly with old stellar populations, 
observed at substantial lookback times, and so one might expect it 
to be difficult to explain such galaxies in a cosmological model in 
which structure grows in a bottom-up fashion through gravitational 
instability. 
Indeed, reproducing the observed abundance of EROs posed a long standing 
challenge to hierarchical galaxy formation models (Smith et~al. 2002). 
Recent theoretical studies have related this problem to the modelling of the suppression of gas cooling in massive haloes as a result of AGN heating (Gonzalez-Perez et~al. 2009) and the need for including the contribution of stars in the thermally-pulsating asymptotic giant branch phase (Fontanot \& Monaco 2010; Henriques et al. 2011).

Using the {\tt GALFORM} semi-analytical model of galaxy formation 
introduced by Cole et~al. (2000), Gonzalez-Perez et~al. (2009) examined the 
predictions for EROs in two published models, those of Baugh et~al. (2005) 
and Bower et~al. (2006). 

The Baugh et~al (2005). model underpredicts the abundance 
of EROs by more than an order of magnitude. This is due in part to the 
long timescale adopted for merger-driven starbursts in this model, which 
means that some residual star formation may be ongoing at a significant time 
after the start of the burst, and because of the top-heavy stellar initial 
mass function (IMF) assumed in starbursts, which, for an old 
stellar population, leads to less K-band light per unit mass of stars 
formed compared with a solar neighbourhood IMF. 
Both of these effects will lead to galaxies 
having bluer optical - near infrared colours, which will tend to move 
them out of the ERO colour selection. 

The Bower et~al. (2006) model, as Gonzalez-Perez et~al. (2009)
demonstrated for the first time, gives a very good match to the 
observed ERO counts. The inclusion of AGN feedback in the Bower et~al. (2006) 
model was one of the main reasons behind this success. Overall, the  
properties of EROs predicted in the Bower et~al. (2006) model show good 
agreement with observations. Most {\tt GALFORM} EROs are quiescent 
galaxies rather than dusty starbursts, 
and have stellar masses $>10^{11} M_{\odot}$ at $z>1$ 
(Gonzalez-Perez et al. 2009). 
Gonzalez-Perez et~al. (2011) broadened the examination of the properties 
of EROs in the Bower et~al. (2006) model to include their clustering. 

In this section, we return to the testing of the predicted clustering 
of EROs in {\tt GALFORM} started by Gonzalez-Perez et~al. (2011). We begin  
by comparing the predicted angular clustering of EROs with the new 
observational measurements presented in this paper (\S~5.1). 
We then discuss the interpretation 
of this clustering in terms of fitted HODs, using the parametric form 
for the HOD given in Eqs.~5, 6, 7 (\S~5.2). {\tt GALFORM} makes 
a direct prediction of the form of the HOD, and we compare   
this {\it intrinsic} HOD with the fitted HOD in \S~5.3.  
The comparison of the model predictions with the observational 
results is discussed in \S~5.4.

\subsection{Galaxy angular clustering comparison} 

% figure : comparison with GALFORM
\begin{figure}
\includegraphics[width=8cm]{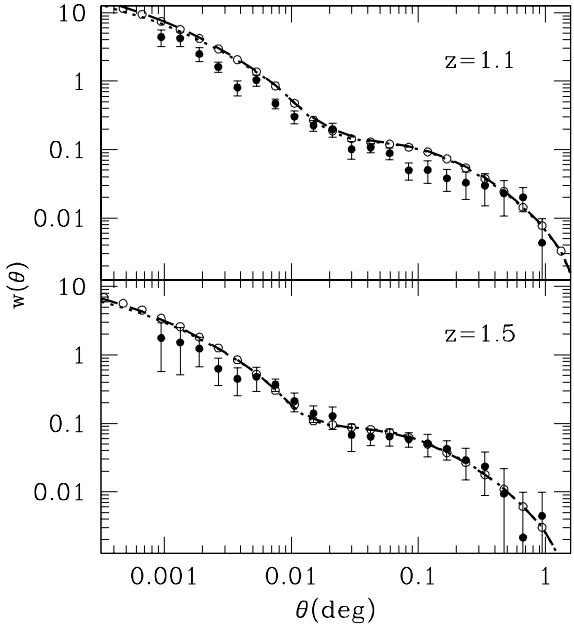}
\caption{The measured angular correlation function of EROs with $M_{K} <-23$ 
at different redshifts compared with predictions from {\tt GALFORM}. 
The upper panel shows the clustering at $z=1.1$ and the lower panel 
that at $z=1.5$. In each case, the filled circles with error bars show 
our new observational measurement of the angular correlation function. 
The open circles show the corresponding predictions from {\tt GALFORM}. 
These are obtained from the HOD output directly by {\tt GALFORM}. The 
lines show attempts to reproduce the clustering predicted by 
{\tt GALFORM}, using different parametric forms for the HOD. 
The dotted line shows the best fit obtained using a standard form 
for the HOD, given by Eqs. 5, 6 \& 7, in which the mean number of 
central galaxies per halo is assumed to reach unity. The dashed line 
shows the angular correlation function resulting from the best 
fit adopting a modified form for the HOD of central galaxies, given 
by Eq.~15 and explained in \S~5.3. In this case, the mean number 
of central galaxies is a HOD parameter. This model gives a better reproduction 
of the {\tt GALFORM} HOD, but gives indistinguishable results for the 
angular correlation function.}
\end{figure}

We start with Fig.~7 in which we compare our new observational 
estimate of the angular correlation function of EROs with 
$(i-K)_{AB}>2.45$ and $M_{K}<-23$ (filled circles) with the 
clustering predicted by {\tt GALFORM} for such galaxies (open circles). 
The model galaxies were extracted with the same colour cut and an apparent 
magnitude\footnote{The magnitudes are $K_{AB}=20.7$ and $21.0$ for $z=1.1$ and $1.5$, respectively.} 
corresponding $M_{K}=-23$ at $z=1.1$ and 1.5.
The theoretical clustering is obtained from the ERO HOD output directly by 
{\tt GALFORM} at a given redshift, along with the predicted redshift distributions. 
We note that this may be different from previous works about the clustering based on 
{\tt GALFORM} and Gonzalez-Perez et al. (2011).
The lines in Fig.~7 show the angular correlation function 
obtained by fitting different parametric forms for the HOD to 
the clustering predicted by {\tt GALFORM} and will be discussed further 
in \S~5.3.

This method of deriving the angular correlation function from 
{\tt GALFORM} does not readily yield an appropriate estimate of the 
error on the angular clustering, as the effective volume considered, the 
volume of the Millennium N-body simulation, is much larger than 
that covered by the observational data. Instead, we 
take the fractional errors inferred on the observational estimate 
of the correlation function, and apply these to the {\tt GALFORM} 
predictions (note this are not plotted in Fig.~7). 
In the future it will be possible to extract exactly 
the same volume sample from the models, using the lightcone mock 
catalogue building capability developed by Merson et~al. (2013). 

The upper panel of Fig.~7 shows that at $z=1.1$ the {\tt GALFORM} 
predictions show slightly stronger angular clustering than is observed, with 
some tension at the 2--3$\sigma$ level (see \S~5.2). 
On large scales, the observational estimate may be affected by sampling 
variance. On the other hand, the discrepancy at small 
angular separations ($\theta<0.02$) can be interpreted as the model 
predicting a larger 1-halo term than is suggested by the observations. 
This in turn implies that the model predicts that too many EROs are 
satellite galaxies at this redshift. 

The lower panel of Fig.~7 shows the angular clustering at $z=1.5$. 
In this case, the predictions agree remarkably well with the new 
observational estimate on large scales. The larger volume covered at $z=1.5$ 
than at $z=1.1$ reduces the effect of cosmic variance. However, 
the discrepancy on small scales also implies a larger satellite 
fraction in the models compared with the observations, in the same 
sense as that suggested at $z=1.1$.

There are several factors which could be responsible for 
the discrepancy between the observed and predicted clustering. 
Firstly, as EROs are unusual objects drawn from the tail of 
luminosity and colour distributions, the precise colour 
criterion applied to construct the ERO samples can have a 
substantial impact on the predictions. Small differences 
between the predicted and observed colour distributions 
can result in large variations in the number of galaxies selected 
and their clustering. Gonzalez-Perez et~al. (2011) noted that the agreement 
between theory and observations was greatly improved on perturbing the colour 
cut used to select EROs in the model. 
Secondly, the semi-analytical modeling predicts too many red satellites resulting in a 
higher clustering amplitude in the 1-halo term. As already
mentioned, most EROs predicted by {\tt GALFORM} are quiescent galaxies
(Gonzalez-Perez et al. 2009). In addition, the clustering amplitude of
quiescent EROs is signicantly higher than dusty, star-forming EROs on small scales in the model 
(Gonzalez-Perez et al. 2011).  However observational results show a smaller discrepancy than this 
model prediction (Miyazaki et al. 2003; Kim et al. 2011a). Hence, it may be necessary to amend 
the prescription for satellite formation in the models
(Kim et al. 2009; Kimm et al. 2009). This issue will be addressed in more detail in \S~5.4.
Thirdly, even for the large number of EROs used in our analysis, 
the relatively small area of the fields (a combined areas of just over 
$5$ square degrees) means that the clustering estimates 
are susceptible to sampling variance. This effect is taken into account 
to some extent in the Jackkinfe errors plotted on our measurements. However, 
Norberg et~al. (2009) demonstrate that an internal estimate of the 
error such as that obtained using Jackknife resampling can still 
vary in amplitude between different realisations of the data, 
particularly if the intrinsic clustering is strong, as is the case with EROs. 

\subsection{Comparison of HODs fitted to observations and GALFORM} 

% table : fitted parameters
\begin{table*}
\caption{The best fitting HOD parameters to the observed angular 
clustering (upper two rows) and the clustering predicted by {\tt GALFORM} 
(lower two rows), assuming the parametric form of Zheng et~al. (2005). 
In both cases, we adopt the cosmology used in the Millennium N-body 
simulation of Springel et~al. (2005). Column 1 gives the redshift of 
the sample; columns 2-5 give the best fitting HOD parameters assuming 
the functional form given by Eqs. 5-7; the remaining columns show 
quantities derived from the HOD fits: column 6 gives the effective bias of 
the EROs (Eq.~13), column 7 lists the effective mass of the haloes which host 
EROs (Eq.~12) and finally column 8 gives the fraction of EROs that are 
satellites according to the fit (Eq.~14).  
}
\centering
\begin{scriptsize}
\begin{tabular}{cccccccc}\\ \hline
median $z$ & $\sigma_{\rm cut}$  & log$(M_{cut}/h^{-1} M_{\odot})$ & log$(M_{0}/h^{-1} M_{\odot})$ & $\alpha$ & $b_{\rm g}$ & log$(M_{\rm eff}/h^{-1} M_{\odot})$ & $f_{\rm sat}$\\
\hline
fit to observations &&&&&&& \\
1.1 & $0.26^{+0.05}_{-0.09}$ &  $12.949^{+0.055}_{-0.045}$ & $14.208^{+0.030}_{-0.025}$ & $1.00^{+0.03}_{-0.03}$ & $2.46^{+0.08}_{-0.07}$ & $13.335^{+0.033}_{-0.027}$ & $0.064^{+0.022}_{-0.018}$\\
1.5 & $0.22^{+0.06}_{-0.07}$ &  $12.927^{+0.086}_{-0.029}$ & $13.944^{+0.023}_{-0.025}$ & $1.82^{+0.11}_{-0.11}$ & $3.12^{+0.09}_{-0.15}$ & $13.278^{+0.037}_{-0.048}$ & $0.040^{+0.014}_{-0.013}$\\
fit to {\tt GALFORM} &&&&&&& \\
1.1 & $0.51^{+0.03}_{-0.03}$ &  $12.588^{+0.041}_{-0.041}$ & $13.462^{+0.021}_{-0.018}$ & $0.79^{+0.04}_{-0.05}$ & $1.92^{+0.04}_{-0.04}$ & $13.059^{+0.033}_{-0.026}$ & $0.160^{+0.027}_{-0.023}$\\
1.5 & $0.61^{+0.03}_{-0.03}$ &  $12.649^{+0.046}_{-0.045}$ & $13.702^{+0.048}_{-0.043}$ & $0.60^{+0.03}_{-0.03}$ & $2.20^{+0.06}_{-0.07}$ & $12.852^{+0.029}_{-0.032}$ & $0.103^{+0.031}_{-0.029}$\\
\hline
\end{tabular}
\end{scriptsize}
\end{table*}

We now compare the HODs derived by fitting the parametric form 
of Zheng et~al. (2005) to the observed clustering and to the clustering 
predicted by the model. To allow a meaningful comparison, 
we perform the HOD analysis of the observed clustering again adopting 
the background cosmology used in the Bower et~al. (2006) model, 
which matches that used in the Millennium N-body simulation of 
Springel et~al. (2005)\footnote{The cosmological parameters used in the Millennium Simulation are 
$\Omega_{\rm m}=$0.25, $\Omega_{\Lambda}=$ 0.75, $\sigma_{8}=0.9$ and $H_{0}=$ 73 
km s$^{-1}$ Mpc$^{-1}$.}. 
%When reanalysing the observed clustering in this way, we will naturally 
%recover somewhat different best fitting HOD parameters compared with our 
%earlier results (Table~2) which assumed a different cosmology.  

The results of this exercise are shown by the dotted lines in Fig.~7, 
which show the angular clustering obtained using the parametric form 
for the HOD given by Eqs. 5, 6 \& 7,  with parameters chosen to give the best fit to the theoretical angular clustering obtained from the {\tt GALFORM} predicted HOD. Table~3 lists the best fitting HOD 
parameters and some derived quantities. The effective bias factors and 
host halo masses deduced from the best fitting HOD to the 
observed clustering are higher than those derived from the 
fits to the predicted clustering. The effective masses are a factor 
of $\sim 1.9$ higher at $z=1.1$ and a factor $\sim 2.7$ higher at $z=1.5$ 
in the fits to the observations compared with the fits to the model.
Another discrepancy between models and observations regards 
the fraction of EROs that are satellites, which is $\approx 2.5$ times 
higher in the model fits than in the fits to the measured clustering. 

The main panels of Fig. 8 display the best fit central HODs to the angular 
clustering of EROs in the observations (solid line) and 
the {\tt GALFORM} predictions (dotted line) at $z=1.1$ (upper)
and $z=1.5$ (lower). The insets of Fig. 8 show the HODs for all galaxies, 
combining the HODs of central and satellite EROs. 
The best fit HOD to the predicted clustering extends to 
lower mass haloes than the fit to the observed clustering. 
In addition, satellites in the model also reside in less massive haloes 
than the observations. 
These departures may explain the higher halo masses and 
lower satellite fractions in the observations than in the model. 
%Comparing them to the HOD output from {\tt GALFORM} (dashed line in Fig.~8), 
%they show different features. 
The form of the HOD output by {\tt GALFORM} will be discussed in next subsection.

%The failure of the model to match the measured clustering on scales 
%corresponding to the 2-halo term implies that {\tt GALFORM} is placing 
%EROs into haloes that are less massive than is implied by the observations. 
%The main panels of Fig. 8 display the best fit HODs of central EROs for 
%observed (solid line) and predicted (dotted line) clustering at $z=1.1$ (upper) 
%and $z=1.5$ (lower). The insets of Fig. 8 show the HODs composed of central 
%and satellite EROs. 
%The 1-halo term is made up of 
%satellite-satellite and satellite-central pairs within the same 
%host halo. Because {\tt GALFORM} is putting galaxies into haloes 
%that are less massive than is suggested by the measured clustering, 
%the transition from the 2-halo to 1-halo term occurs at a different 
%scale than it does in the data (as the virial radius of the host 
%haloes is smaller than it should be). The fact that the predicted 
%and measured clustering agree on small scales does not guarantee 
%that the fraction of satellites predicted by {\tt GALFORM} matches 
%that inferred from the observations, if, as we find, there is 
%disagreement between the model and observations on large scales. 

\subsection{The intrinsic form of the HOD predicted by GALFORM}

% table : halo masses and r_0 from GALFORM
\begin{table*}
\caption{Basic properties of the ERO samples predicted by 
{\tt GALFORM}. Column 1 gives the redshift of the 
ERO sample. Column 2 gives the number density 
of EROs. Columns 3-6 give various measures of the 
distribution of dark matter haloes which host EROs: 
the logarithm of the median host halo mass, the 10 and 90 percentiles 
of the distribution (weighted by the number of EROs hosted by each halo), 
the effective mass (defined as in Eq. 12). 
Column 7 gives the fraction of EROs that are satellite galaxies 
(Eq.~14). Note that these quantities are computed directly from 
the model output, rather than from a fitted HOD.  
}
\centering
\begin{scriptsize}
\begin{tabular}{ccccccc}\\ \hline
redshift 
& $n_{\rm g}$                
& $\log (M_{50\%}/ h^{-1} M_{\odot} )$ 
& $\log (M_{10\%}/ h^{-1} M_{\odot} )$ 
& $\log (M_{90\%}/ h^{-1} M_{\odot} )$ 
& $\log (M_{\rm eff}/ h^{-1} M_{\odot} )$ 
& $f_{\rm sat}$\\
         
& $(10^{-4} h^{3} {\rm Mpc}^{-3})$ 
&    
&                
&        
&
&                        
\\
\hline
1.1 & 11.8 & 12.572 & 11.918 & 13.523 & 13.105 & 0.25\\
1.5 &  8.8 & 12.320 & 11.784 & 13.199 & 12.903 & 0.19\\
\hline
\end{tabular}
\end{scriptsize}
\end{table*}

The galaxy HOD is a prediction of {\tt GALFORM} and not an 
input. {\tt GALFORM} models the physics of the baryonic component 
of the Universe to predict the number of galaxies per halo and their properties. 
The HOD is extracted by applying the observational selection to 
the model output and simply counting the number of galaxies 
which are retained in each dark matter halo, distinguishing 
between the central galaxy and its satellites.   
The HOD and the quantities derived from it can be extracted 
directly from the model, without having to go through the intermediate step 
of fitting a parametric form for the HOD to the predicted clustering. As we 
will see later on in this subsection, the form of the actual HOD output by 
{\tt GALFORM} can be different from the standard parametrization we 
have adopted so far. 

We begin by listing in Table~4 some basic properties predicted by {\tt GALFORM} 
for the EROs and their host haloes. 
The effective host halo mass defined by Eq. 12 with the direct output HOD 
of {\tt GALFORM} is in very close agreement with that derived 
from the HOD fit to the predicted angular clustering listed in Table 3. 
This means that EROs predicted by {\tt GALFORM} are in less massive haloes 
than the observations. The major differences between the model prediction 
and observations are number density and satellite fraction.
The predicted number density 
of EROs is much larger than the observational estimate in the previous section. 
Also the fraction of EROs that are satellites from the direct model prediction 
is around 4--5 times higher than 
the fraction obtained from the best fitting HOD to the observation. 

% figure : Halo occupation distribution
\begin{figure}
\includegraphics[width=8cm]{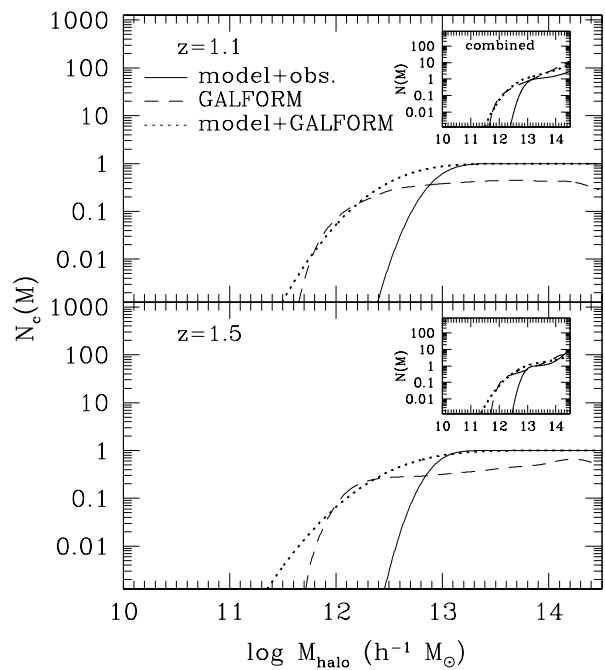}
\caption{
The Halo Occupation Distributions of EROs brighter than 
$M_{K} = -23$ with  $(i-K)_{AB}>2.45$ EROs at $z=1.1$ (top panel) 
and $z=1.5$ (bottom panel). The central galaxy HOD predicted 
directly by {\tt GALFORM} is shown by the dashed lines in the 
main panels. The dotted lines show the HOD fitted to the angular 
clustering predicted by {\tt GALFORM}, when using the 
parametrization of Eqs.~5-7. The solid line shows the central 
galaxy HOD fitted to the observational estimate of the angular 
clustering using the same form. The inset shows the combined HOD of 
central and satellite EROs, with the lines retaining the 
meaning they have in the main panel. 
}
\label{fig:GALFORMHOD}
\end{figure}

% figure : Halo occupation distribution
\begin{figure}
\includegraphics[width=8cm]{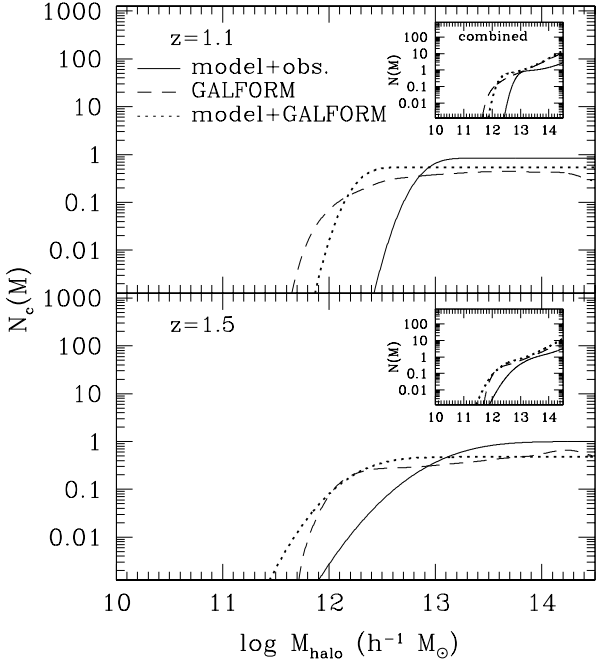}
\caption{
Halo occupation distributions of EROs satisfying 
$M_{K}< -23$ and $(i-K)_{AB}>2.45$ at different 
redshifts with a modified $N_{\rm c}(M)$, as given by Eq.~15. 
The line styles are the same as those used in Fig.~8.
}
\end{figure}

We plot the HOD predicted directly by {\tt GALFORM} in 
Fig.~\ref{fig:GALFORMHOD} (dashed line). 
We compare this with the parameterised  
form for the HOD proposed by Zheng et~al. (2005), which was 
motivated by earlier simulations of galaxy formation which did 
not include AGN feedback. In the Zheng et~al. (2005) framework, the 
HOD of central galaxies is assumed to reach unity, i.e. above
 some halo mass, every central galaxy is assumed to meet the 
selection criteria of the sample. The parametric HOD 
(dotted line) plotted in Fig.~\ref{fig:GALFORMHOD} is the best 
fit to the angular clustering predicted by {\tt GALFORM}. 

The predicted HOD can differ substantially from the canonical 
form typically assumed for the HOD as shown in Fig.~\ref{fig:GALFORMHOD}
and remarked upon by Contreras et~al. (2013) in their comparison of the HOD 
predicted in different semi-analytical models.  
The deviation is largely driven by AGN feedback, which shuts down gas cooling in 
massive haloes in {\tt GALFORM}. In general, the onset of AGN feedback 
above some halo mass alters how galaxy properties vary with 
halo mass. This could be manifest as a dramatic break in the relation between 
the galaxy property and halo mass, as in the case of cold gas 
mass (Kim et~al. 2011b), or as a change in the slope and 
scatter of the relation, as in the case of $K$-band luminosity (Gonzalez-Perez 
et~al. 2011). 
With AGN feedback, the most massive galaxy in a sample may no longer be 
in the most massive dark matter halo due to the increased scatter in the 
correlation between galaxy properties and halo mass. 

Fig.~\ref{fig:GALFORMHOD} shows the HOD of central galaxies 
that are EROs at $z=1.1$ (upper) and $1.5$ (lower), respectively. 
The predicted HOD of central EROs differs from the observationally 
inferred in terms of the transition from zero galaxies per halo and 
also in the number of central galaxies which are EROs. 
At both redshifts, the mean number of centrals is below unity 
for the predicted HOD. In addition the fitted HOD shows a 
smoother shape than the {\tt GALFORM} prediction. These features 
emphasize the importance of the form adopted for the HOD when 
interpreting the results of halo modeling.

Motivated by the comparison between the predicted and fitted HODs 
in Fig.~\ref{fig:GALFORMHOD}, we explore fitting a modified parametric 
form for the HOD to the angular clustering predicted by 
{\tt GALFORM}. The new form allows the mean number of central galaxies 
as a function of halo mass, $N_{\rm c} (M)$, to vary rather than forcing 
it to be unity: 
\begin{equation}
N_{\rm c}(M) = 0.5 C_{\rm amp} \left[ 1 + {\rm erf}\left(\frac{{\log_{10}} (M/M_{\rm cut})}{\sigma_{\rm cut}}\right)\right], 
\end{equation}
\noindent where $C_{\rm amp}$ sets the maximum mean number of 
central EROs and is allowed to take on values $\leq 1$. 
Since we have introduced an additional free parameter with this 
formulation of the HOD, we fix the exponent of the 
power-law slope ($\alpha$) for satellites to unity, as derived 
in previous work for SDSS galaxies (Zehavi et al. 2011) and 
the {\tt GALFORM} prediction. 
The best fit angular correlation function using this revised 
parametric form for the HOD is shown in Fig.~7 by the dashed lines. 
Fig.~9 shows the best fitting HODs to the measured and predicted 
angular clustering using this modified form. 
The line styles are the same as used in Fig.~8. 
The parametric HOD of central EROs which gives the best fit 
to the predicted clustering is now closer to the direct HOD 
prediction of {\tt GALFORM}. However, the best fit HOD to the {\tt GALFORM} clustering  
predictions still extends to lower halo masses than the best fit to the observed 
angular clustering. It is interesting to note  
when fitted to the observed ERO clustering (solid lines) at $z=1.1$ 
$N_{\rm C}$ may be possible to be below unity, with a best fitting value of 
$C_{\rm amp}=0.84^{+0.12}_{-0.13}$, but does so at $z=1.5$, with 
$C_{\rm amp} = 1^{+0}_{-0.06}$. These values may indicate the onset 
of AGN feedback between these two redshifts. 
However, this interpretation is complicated by the fact that 
EROs do not correspond readily to a mass limited sample, which may 
lead to some central galaxies being omitted as they have the 
different colours, offering an alternative explanation for 
a mean number of central galaxies that is less than unity. 
All deduced parameters are listed in Table~5. 

The modified HOD fits the HOD of central EROs in {\tt GALFORM} 
much better at both redshifts than was possible using the 
Zheng et~al. (2005) HOD. This emphasizes the importance of using 
a realistic form for the HOD form to derive robust halo properties 
from the galaxy distribution. 
 
% table : fitted parameters
\begin{table*}
\caption{The best fitting HOD parameters and derived quantities using the 
observed clustering (upper two rows) and the clustering predicted by 
{\tt GALFORM} with the modified HOD for central EROs as Eq.~15. In this case, 
we fix the slope ($\alpha$) for the HOD of satellites as 1. Therefore column 5 is 
the maximum mean number ($C_{amp}$) of central galaxies in Eq.~15. The other columns 
are the same as Table~3.}
\centering
\begin{scriptsize}
\begin{tabular}{cccccccc}\\ \hline
median $z$ & $\sigma_{\rm cut}$  & log$(M_{\rm cut}/h^{-1} M_{\odot})$ & log$(M_{0}/h^{-1} M_{\odot})$ & $C_{amp}$ & $b_{\rm g}$ & log$(M_{\rm eff}/h^{-1} M_{\odot})$ & $f_{\rm sat}$\\
\hline
fit to observations &&&&&&& \\
1.1 & $0.22^{+0.07}_{-0.05}$ &  $12.884^{+0.107}_{-0.125}$ & $14.197^{+0.032}_{-0.032}$ & $0.84^{+0.12}_{-0.13}$ & $2.42^{+0.12}_{-0.19}$ & $13.310^{+0.059}_{-0.085}$ & $0.073^{+0.028}_{-0.018}$\\
1.5 & $0.58^{+0.03}_{-0.05}$ &  $13.140^{+0.004}_{-0.116}$ & $14.121^{+0.049}_{-0.055}$ & $1.00^{+0.00}_{-0.06}$ & $2.77^{+0.03}_{-0.18}$ & $13.158^{+0.008}_{-0.072}$ & $0.038^{+0.025}_{-0.008}$\\
fit to {\tt GALFORM} &&&&&&& \\
1.1 & $0.20^{+0.12}_{-0.08}$ &  $12.271^{+0.112}_{-0.057}$ & $13.342^{+0.015}_{-0.015}$ & $0.54^{+0.02}_{-0.02}$ & $1.96^{+0.05}_{-0.08}$ & $13.117^{+0.027}_{-0.034}$ & $0.260^{+0.027}_{-0.036}$\\
1.5 & $0.43^{+0.07}_{-0.09}$ &  $12.293^{+0.201}_{-0.140}$ & $13.403^{+0.028}_{-0.031}$ & $0.47^{+0.06}_{-0.07}$ & $2.17^{+0.15}_{-0.15}$ & $12.871^{+0.073}_{-0.067}$ & $0.181^{+0.060}_{-0.051}$\\
\hline
\end{tabular}
\end{scriptsize}
\end{table*}

\subsection{What went wrong?}

The comparisons above point to {\tt GALFORM} predicting that,   
overall, EROs are found in less massive haloes and that more EROs are 
satellites compared with the conclusions reached by fitting HOD models 
to the observed clustering.  

Both problems could be solved by changing the treatment of 
cooling gas in {\tt GALFORM}. Gonzalez-Perez et~al. (2011) 
looked at the predictions of a different semi-analytical model, that of 
Font et~al. (2008). The Font et~al. model includes partial stripping 
of the hot gas halo from satellite galaxies, whereas the default 
assumption is that all of the hot gas is stripped from a galaxy once 
it becomes a satellite within a larger halo. In the Font et~al. model, 
depending on the orbit of the satellite and the ram pressure that it 
experiences, some hot gas may be retained and can cool onto the galaxy even 
after it has become a satellite. This change to the model makes satellites 
bluer, by permitting more star formation to take place. However, Font et~al. 
also invoked a factor of two increase in the stellar yield without changing 
the IMF, which leads to redder galaxy colours. Gonzalez-Perez et~al. 
report that the Font et al model produces stronger clustering for EROs than 
the model of Bower et~al. but also leads to more EROs than are observed.  

Additionally, Gonzalez-Perez et~al. (2009) mentioned that the Bower et al. 
(2006) model quenched the star formation of massive galaxies too efficiently by 
the AGN feedback, based on the predicted redshift distribution of passive galaxies. 
Zheng et al. (2009) also pointed out that this model predicts more red galaxies 
in a few times $10^{12} h^{-1} M_{\odot}$ haloes than observations but less in 
more massive haloes, comparing the HODs of observed and predicted LRGs. Therefore 
it may be possible that {\tt GALFORM} predicts more red central galaxies such as EROs 
and LRGs in lower mass haloes.

These comparison illustrate the potential of our new clustering 
results to constrain the modeling of different elements of the 
physics of galaxy formation. 

\section{Conclusion}

Recently wide and deep near-IR surveys have made it possible to select $z>1$ 
galaxies effectively. In this study we have used a near-IR dataset from the 
UKIDSS DXS and optical datasets from Pan-STARRS PS1 and Subaru to investigate 
the clustering of EROs and the halo properties hosting them. 
The main results can be summarised as follows;

\begin{enumerate}

\item $(i-K)$ colour cuts were applied to extract EROs from 
the 3.88 deg$^{2}$ DXS/Subaru and the 5.33 deg$^{2}$ DXS/PS1 catalogues, 
respectively. We detected 17\,250 EROs from DXS/PS1 and 23\,916 EROs from 
DXS/Subaru. The number counts of EROs agree well with previous 
studies. The photometric redshifts of galaxies in DXS/PS1 were measured from 
$grizJK$ and SWIRE IRAC colours. These EROs were split into subsamples of different 
photometric redshift and brighter than a fixed absolute magnitude ($M_{K}<-23$). 

\item The angular correlation functions of EROs were measured with several 
colour and magnitude thresholds from the DXS/Subaru sample. All these correlation 
functions 
showed a clear break at $\sim 0.02^{\circ}$ which implies that the angular correlation 
function of EROs cannot be described by a single power-law. Furthermore, 
redder or brighter EROs showed higher amplitudes than bluer or fainter ones. 
The correlation functions from the DXS/PS1 and the DXS/Subaru samples with the 
same 
criteria also showed good agreement. The correlation functions at different 
redshifts had similar amplitudes on large scales, indicating a 
higher bias at higher redshift.

\item A standard halo model was fitted to the observed angular 
correlation 
of EROs and matches well. The biases for EROs range between 2.7 and 3.5, and the average dark 
matter halo mass hosting EROs is over $10^{13} h^{-1} M_{\odot}$. EROs at 
higher redshifts are more biased and located in slightly less massive 
dark matter haloes than at lower redshift. Also the ERO satellite fraction 
decreases with increasing redshift. The different fraction of old, passive EROs 
at different redshifts may affect the properties.
In addition, the overall halo properties for EROs are 
consistent with $10^{11.0} M_{\odot}<M_{*}<10^{11.5} M_{\odot}$ galaxies.

\item The predicted angular correlation function of EROs from the 
{\tt GALFORM} semi-analytic model showed good agreement with the observed 
correlation function. Comparing the halo model 
for observed EROs to the {\tt GALFORM} predictions, we found that the EROs 
predicted by {\tt GALFORM} present too high a fraction of satellite 
galaxies or too many galaxies in less massive haloes. Finally, 
we stress that the results from the HOD frame work must be interpreted with care due 
to the effect of AGN feedback and that additional refinements are necessary in  
future semi-analytical models to improve the modelling of the physics of galaxy 
formation.

\end{enumerate}

Our clustering results are 
derived from the large solid angle survey currently available. 
Nevertheless, the effects of cosmic variance dominate at large scales, so 
substantial improvements could be obtained in the measurements from even 
larger surveys. In the near future, the completed UKIDSS and VISTA surveys will 
be important for making further progress on studying galaxy evolution at $z>1$.
Moreover the combination of near-IR surveys and improved 
optical surveys such as Pan-STARRS, Hyper Suprime Camera on Subaru and LSST 
will have a dramatic impact.

\section*{Acknowledgments}

Authors thank referee for comments improving the paper. 
This work is based on 
the data from UKIRT Infrared Deep Sky Survey. We are grateful to UKIDSS team, 
the staff in UKIRT, Cambridge Astronomical Survey Unit and Wide Field 
Astronomy Unit in Edinburgh. The United Kingdom Infrared Telescope is run by 
the Joint Astronomy Centre on behalf of the Science and Technology Facilities 
Council of the U.K.
The PS1 Surveys have been made possible through contributions of the Institute 
for Astronomy, the University of Hawaii, the Pan-STARRS Project Office, the 
Max-Planck Society and its participating institutes, the Max-Planck Institute for 
Astronomy, Heidelberg and the Max-Planck Institute for Extraterrestrial Physics, 
Garching, The Johns Hopkins University, Durham University, the University of 
Edinburgh, Queen's University Belfast, the Harvard-Smithsonian Center for Astrophysics, 
the Las Cumbres Observatory Global Telescope Network, Incorporated, the National Central 
University of Taiwan, and the National Aeronautics and Space Administration under 
grant no. NNX08AR22G issued through the Planetary Science Division of the NASA 
Science Mission Directorate.
This work is also partially based on data collected at Subaru Telescope, which is operated 
by the National Astronomical Observatory of Japan. 
{\tt GALFORM} was run on the ICC Cosmological machine, which is part of the 
DiRAC Facility jointly
funded by STFC, the Large Facilities Capital Fund of BIS, and
Durham University.
JWK acknowledges the support from the Creative Research Initiative program, 
No. 2008-0060544, of the National Research Foundation of Korea (NRF) 
funded by the Korea government (MSIP).
ACE acknowledges support from STFC grant ST/I001573/1. 
VGP, CMB and CGL acknowledge support from STFC grant ST/F001166/1.

\label{lastpage}


\begin{thebibliography}{99}
\bibitem[\protect\citeauthoryear{Almeidaetal}{2008}]{b1} Almeida C., Baugh C. M., 
Wake D. A., Lacey C. G., Benson A. J., Bower R. G., Pimbblet K., 2008, MNRAS, 386, 2145
\bibitem[\protect\citeauthoryear{Barberetal}{2007}]{b1} Barber T., Meiksin A., 
Murphy T., 2007, MNRAS, 377, 787
\bibitem[\protect\citeauthoryear{Baugh}{2006}]{b2} Baugh C. M., 2006, Reports
on Progress in Physics, 69, 3101
\bibitem[\protect\citeauthoryear{Baughetal}{2005}]{b2} Baugh C. M., Lacey C. 
G., Frenk C. S., Granato G. L., Silva L., Bressan A., Benson A. J., Cole S., 
2005, MNRAS, 356, 1191
\bibitem[\protect\citeauthoryear{Bensonetal}{2000}]{b1} Benson A. J., Cole 
S., Frenk C. S., Baugh C. M., Lacey C. G., 2000, MNRAS, 311, 793
\bibitem[\protect\citeauthoryear{BerlindandWeiberg}{2002}]{b1} Berlind A. A., 
Weinberg D. H., 2002, ApJ, 575, 587
\bibitem[\protect\citeauthoryear{Bertaetal}{2007}]{b1} Berta S. et al., 2007, 
A\&A, 467, 565
\bibitem[\protect\citeauthoryear{Bertin \& Arnouts}{1996}]{b1} Bertin 
E., Arnouts S., 1996, A\&AS, 117, 393
\bibitem[\protect\citeauthoryear{Bertinetal}{1996}]{b2} Bertin E., Mellier Y., 
Radovich M., Missonier G., Didelon P., Morin B., 2002, in Bohlender D. A., 
Durand, D., Handley T. H., eds, ASP Conf. Ser., Vol. 281, Astronomical Data 
Analysis Software and Systems XI. Astron. Soc. Pac., San Francisco, p. 228
\bibitem[\protect\citeauthoryear{Blakeetal}{2008}]{b5} Blake C., Collister A., 
Lahav O., 2008, MNRAS, 385, 1257
\bibitem[\protect\citeauthoryear{Boweretal}{2006}]{b5} Bower R. G., Benson A. 
J., Malbon R., Helly J. C., Frenk C. S., Baugh C. M., Cole S., Lacey C. G., 
2006, MNRAS, 370, 645
\bibitem[\protect\citeauthoryear{Brammeretal}{2008}]{b5} Brammer G. B., 
van Dokkum P. G., Coppi P., 2008, ApJ, 686, 1503
\bibitem[\protect\citeauthoryear{Brammeretal}{2009}]{b8} Brammer G. B.
et al., 2009, ApJ, 706, 173
\bibitem[\protect\citeauthoryear{Brownetal}{2005}]{b9} Brown M. J. I., Jannuzi
B. T., Dey A., Tiede G. P., 2005, ApJ, 621, 41
\bibitem[\protect\citeauthoryear{Bruzual}{2003}]{b9} Bruzual G., Charlot S., 
2003, MNRAS, 344, 1000
\bibitem[\protect\citeauthoryear{Casalietal}{2007}]{b10} Casali M. et al.,
2007, A\&A, 467, 777
\bibitem[\protect\citeauthoryear{Chabrier}{2003}]{b11} Chabrier, G. 2003, PASP, 
115, 763
\bibitem[\protect\citeauthoryear{Cimattietal}{2002}]{b11} Cimatti A. et al.,
2002, A\&A, 381, L68
\bibitem[\protect\citeauthoryear{Cimattietal}{2003}]{b12} Cimatti A. et al.,
2003, A\&A, 412, L1
\bibitem[\protect\citeauthoryear{Coiletal}{2006}]{b12} Coil A. L., Newman 
J. A., Cooper M. C., Davis M., Faber S. M., Koo D. C., Willmer C. N. A., 
2006, ApJ, 644, 671
\bibitem[\protect\citeauthoryear{Coiletal}{2008}]{b12} Coil A. L. et al., 
2008, ApJ, 672, 153
\bibitem[\protect\citeauthoryear{Coiletal}{2009}]{b12} Coil A. L. et al., 
2009, ApJ, 701, 1484
\bibitem[\protect\citeauthoryear{Coleetal}{2000}]{b12} Cole S., Lacey C. G., 
Baugh C. M., Frenk C. S., 2000, MNRAS, 319, 168
\bibitem[\protect\citeauthoryear{Conseliceetal}{2008}]{b14} Conselice C. J.,
Bundy K., U V., Eisenhardt P., Lotz J., Newman J., 2008, MNRAS, 383, 1366
\bibitem[\protect\citeauthoryear{Contrerasetal}{2013}]{b14} Contreras S., 
Baugh C., Norberg P., Padilla N., 2013, MNRAS, 432, 2717
\bibitem[\protect\citeauthoryear{CooraySheth}{2002}]{b15} Cooray A., Sheth R.,
2002, Physics Reports, 372, 1
\bibitem[\protect\citeauthoryear{Croom05}{2005}]{c99} Croom, S. M., Boyle, B. J.,
 Shanks, T., Smith, R. J., Miller, L., Outram, P. J., Loaring, N. S., Hoyle, F. da Angela, J.,
2005, MNRAS, 356, 415
\bibitem[\protect\citeauthoryear{Daddietal}{2000}]{b17} Daddi E., Cimatti A.,
Pozzetti L., Hoekstra H., R\"{o}ttgering H. J., Renzini A., Zamorani G.,
Mannucci F., 2000, A\&A, 361, 535
\bibitem[\protect\citeauthoryear{Daddietal}{2004}]{b18} Daddi E., Cimatti A.,
Renzini A., Fontana A., Mignoli M., Pozzetti L., Tozzi P., Zamorani G.,
2004, ApJ, 617, 746
\bibitem[\protect\citeauthoryear{Dyeetal}{2006}]{b20} Dye S. et al., 2006,
MNRAS, 372, 1227
\bibitem[\protect\citeauthoryear{Ekeetal}{2004}]{b21} Eke V. R. et al.,
2004, MNRAS, 355, 769
\bibitem[\protect\citeauthoryear{Elstonetal}{1988}]{b22} Elston R., Rieke
G. H., Rieke M. J., 1988, ApJ, 331, 77
%\bibitem[\protect\citeauthoryear{FallandEfstathiou}{1980}]{b5} Fall S. M., 
%Efstathiou G., 1980, MNRAS, 193, 189 
\bibitem[\protect\citeauthoryear{Firthetal}{2002}]{b5} Firth A. E. et al., 
2002, MNRAS, 332, 617
\bibitem[\protect\citeauthoryear{Fontetal}{2008}]{b24} Font A. S. et al., 
2008, MNRAS, 389, 1619
\bibitem[\protect\citeauthoryear{FontanotMonaco}{2010}]{b24} Fontanot F., 
Monaco P., 2010, MNRAS, 405, 705
\bibitem[\protect\citeauthoryear{Foucaudetal}{2007}]{b24} Foucaud S. et al.,
2007, MNRAS, 376, L20
\bibitem[\protect\citeauthoryear{Foucaudetal}{2010}]{b5} Foucaud S., 
Conselice C. J., Hartley W. G., Lane K. P., Bamford S. P., Almaini O., Bundy 
K., 2010, MNRAS, 406, 147
\bibitem[\protect\citeauthoryear{Franxetal}{2003}]{b26} Franx M. et al., 2003,
ApJ, 587, L79
\bibitem[\protect\citeauthoryear{Furusawaetal}{2011}]{b5} Furusawa J., 
Sekiguchi K., Takata T., Furusawa H., Shimasaku K., Simpson C., Akiyama 
M., 2011, ApJ, 727, 111
\bibitem[\protect\citeauthoryear{Georgakakisetal}{2005}]{b31} Georgakakis A., 
Afonso J., Hopkins A. M., Sullivan M., Mobasher B., Cram L. E., 2005, ApJ, 
620, 584
\bibitem[\protect\citeauthoryear{Gonzalez-perezatal}{2009}]{b31} Gonzalez-Perez 
V., Baugh C. M., Lacey C. G., Almeida, C., 2009, MNRAS, 398, 497
\bibitem[\protect\citeauthoryear{Gonzalez-perezatal}{2011}]{b31} Gonzalez-Perez 
V., Baugh C. M., Lacey C. G., Kim J. -W., 2011, MNRAS, 417, 517
\bibitem[\protect\citeauthoryear{Grazianetal}{2006}]{b30} Grazian A. et al.,
2006, A\&A, 453, 507
\bibitem[\protect\citeauthoryear{GrothPeebles}{1977}]{b31} Groth E. J.,
Peebles, P. J. E., 1977, ApJ, 217, 385
\bibitem[\protect\citeauthoryear{GuoWhite}{2009}]{b31} Guo Q., White S. D. M., 
2009, MNRAS, 396, 39
\bibitem[\protect\citeauthoryear{Hartleyetal}{2008}]{b31} Hartley W. G. et al., 
2008, MNRAS, 391, 1301
\bibitem[\protect\citeauthoryear{Hartleyetal}{2010}]{b31} Hartley W. G. et al., 
2010, MNRAS, 407, 1212
\bibitem[\protect\citeauthoryear{Henriquesetal}{2011}]{b31} Henriques B., 
Maraston C., Monaco P., Fontanot F., Menci N., De Lucia G., Tonini C., 2011, 
MNRAS, 415, 3571
%\bibitem[\protect\citeauthoryear{Hempeletal}{2011}]{b31} Hempel A. et al., 
%2011, MNRAS, 414, 2246
\bibitem[\protect\citeauthoryear{Hickoxetal}{2011}]{b31} Hickox R. C. et al., 
2011, ApJ, 731, 117
\bibitem[\protect\citeauthoryear{Jingetal}{1998}]{b5} Jing Y. P., Mo H. J., 
Boerner G., 1998, ApJ, 494, 1
\bibitem[\protect\citeauthoryear{Kaiser}{2002}]{b5} Kaiser N., Pan-STARRS team, 
2002, BAAS, 34, 1304
\bibitem[\protect\citeauthoryear{Kimetal}{2011}]{b5} Kim H. -S., Baugh C. M., 
Benson A. J., Cole S., Frenk C. S., Lacey C. G., Power C., Schneider M., 
2011b, MNRAS, 414, 2367
\bibitem[\protect\citeauthoryear{Kimetal}{2009}]{b5} Kim H. -S., Baugh C. M., 
Cole S., Frenk C. S., Benson A. J., 2009, MNRAS, 400, 1527
\bibitem[\protect\citeauthoryear{Kimetal}{2011}]{b5} Kim J. -W., Edge A. C., 
Wake D. A., Stott, J. P., 2011a, MNRAS, 410, 241 
\bibitem[\protect\citeauthoryear{Kimmetal}{2009}]{b5} Kimm T. et al. 2009, 
MNRAS, 394, 1131
\bibitem[\protect\citeauthoryear{Kongetal}{2006}]{b34} Kong X. et al., 2006,
ApJ, 638, 72
\bibitem[\protect\citeauthoryear{Kongetal}{2009}]{b35} Kong X., Fang G.,
Arimoto N., Wang M., 2009, ApJ, 702, 1458
\bibitem[\protect\citeauthoryear{Landy \& Szalay}{1993}]{b38} Landy S. D.,
Szalay A. S., 1993, ApJ, 412, 64
\bibitem[\protect\citeauthoryear{Laneetal}{2007}]{b38} Lane K. P. et al., 
2007, MNRAS, 379, L25
\bibitem[\protect\citeauthoryear{Lawrenceetal}{2007}]{b40} Lawrence A. et al.,
2007, MNRAS, 379, 1599
\bibitem[\protect\citeauthoryear{Lewisetal}{2000}]{b40} Lewis A., Challinor 
A., Lasenby A., 2000, ApJ, 538, 473
\bibitem[\protect\citeauthoryear{Limber}{1954}]{b38} Limber D. N., 1954, ApJ, 
119, 655
\bibitem[\protect\citeauthoryear{Lonsdaleetal}{2003}]{b5} Lonsdale C. J. et 
al., 2003, PASP, 115, 897
\bibitem[\protect\citeauthoryear{MaanFry}{2000}]{b5} Ma C. -P., Fry J. N.,
2000, ApJ, 543, 503
\bibitem[\protect\citeauthoryear{Matsuokaetal}{2011}]{b5} Matsuoka Y., Masaki 
S., Kawara K., Sugiyama N., 2011, MNRAS, 410, 548
\bibitem[\protect\citeauthoryear{McCarthyetal}{2001}]{b5} McCarthy P. J. 
et al., 2001, ApJ, 560, L131
\bibitem[\protect\citeauthoryear{McCrackenetal}{2010}]{b5} McCracken T. M. 
et al., 2010, ApJ, 708, 202
\bibitem[\protect\citeauthoryear{Mersonetal}{2013}]{b5} 
Merson A. I. et al., 2013, MNRAS, 429, 556
\bibitem[\protect\citeauthoryear{Miyazakietal}{2003}]{b5} Miyazaki M. et al., 
2003, PASJ, 55, 1079
\bibitem[\protect\citeauthoryear{Miyazakietal}{2002}]{b5} Miyazaki S. et al., 
2002, PASJ, 54, 833
\bibitem[\protect\citeauthoryear{MoandWhite}{1996}]{b5} Mo H. J., White S. D. 
M., 1996, MNRAS, 282, 347
\bibitem[\protect\citeauthoryear{MoandWhite}{2002}]{b5} Mo H. J., White S. D. 
M., 2002, MNRAS, 336, 112
\bibitem[\protect\citeauthoryear{MoustakasSomerville}{2002}]{b5} Moustakas L. 
A., Somerville R. S., 2002, ApJ, 577, 1
\bibitem[\protect\citeauthoryear{Moustakasetal}{2004}]{b45} Moustakas L. A. et
al., 2004, ApJ, 600, L131
\bibitem[\protect\citeauthoryear{Myersetal}{2009}]{b45} Myers A. D., White M., 
Ball N. M., 2009, MNRAS, 399, 2279
\bibitem[\protect\citeauthoryear{Navarroetal}{1997}]{b5} Navarro J. F., Frenk 
C. S., White S. D. M., 1997, ApJ, 490, 493
\bibitem[\protect\citeauthoryear{Nikoloudakisetal}{2013}]{b5} Nikoloudakis N., 
Shanks T., Sawangwit U., 2013, MNRAS, 429, 2032
\bibitem[\protect\citeauthoryear{Norbergetal}{2001}]{b5} Norberg P. Baugh C. 
M., Gazta\~{n}aga E., Croton D. J., 2009, MNRAS, 396, 19 
\bibitem[\protect\citeauthoryear{Norbergetal}{2001}]{b5} Norberg P. et al., 
2001, MNRAS, 328, 64
\bibitem[\protect\citeauthoryear{Norbergetal}{2002}]{b5} Norberg P. et al., 
2002, MNRAS, 332, 827
\bibitem[\protect\citeauthoryear{Ouchi}{2004}]{b5} Ouchi M., 2004, ApJ, 611, 660
\bibitem[\protect\citeauthoryear{Palamaraetal}{2013}]{b48} Palamara D. P. 
et al., 2013, ApJ, 764, 31
\bibitem[\protect\citeauthoryear{PeacockandSmith}{2000}]{b48} Peacock J. A., 
Smith R. E., 2000, MNRAS, 318, 1144
\bibitem[\protect\citeauthoryear{Peebles}{1980}]{b48} Peebles P. J. E., 1980,
The Large-Scale Structure of the Universe. Princeton Univ, Press, Princeton,
NJ.
\bibitem[\protect\citeauthoryear{Pozzetti \& Mannucci}{2000}]{b50} Pozzetti L.,
Mannucci F., 2000, MNRAS, 317, L17
\bibitem[\protect\citeauthoryear{QuadriWilliams}{2010}]{b5} Quadri R. F., 
Williams R. J., 2010, ApJ, 725, 794
\bibitem[\protect\citeauthoryear{Quadrietal}{2008}]{b52} Quadri R. F.,
Williams R. J., Lee K., Franx M., van Dokkum P., Brammer G. B., 2008, ApJ, 685,
L1
\bibitem[\protect\citeauthoryear{Rocheetal}{2002}]{b53} Roche N. D., Almaini O.,
Dunlop J., Ivison R. J., Willott C. J., 2002, MNRAS, 337, 1282
\bibitem[\protect\citeauthoryear{Rocheetal}{2003}]{b54} Roche N. D., Dunlop J.,
Almaini O., 2003, MNRAS, 346, 803
\bibitem[\protect\citeauthoryear{Rocheetal}{1999}]{b55} Roche N., Eales S.
A., Hippelein H., Willott C. J., 1999, MNRAS, 306, 538
\bibitem[\protect\citeauthoryear{Rossetal}{2007}]{b55} Ross N. P. et al., 2007, 
MNRAS, 381, 573
\bibitem[\protect\citeauthoryear{RossN2009}{2009}]{b55} 
Ross N. P. et al., 2009, ApJ, 697, 1634
\bibitem[\protect\citeauthoryear{Rossetal}{2009}]{b55} Ross A. J., Brunner R. 
J., 2009, MNRAS, 399, 878
\bibitem[\protect\citeauthoryear{Rossetal}{2010}]{b55} Ross A. J., Percival 
W. J., Brunner R. J., 2010, MNRAS, 407, 420
\bibitem[\protect\citeauthoryear{RowanRobinsonetal}{2008}]{b5} Rowan-Robinson 
M. et al., 2008, MNRAS, 386, 697
\bibitem[\protect\citeauthoryear{Sawangwitetal}{2009}]{b58} Sawangwit U.,
Shanks T., Abdalla F. B., Cannon R. D., Croom S. M., Edge A. C., Ross N. P.,
Wake D. A., 2011, MNRAS, 416, 3033
\bibitem[\protect\citeauthoryear{Sawickietal}{2005}]{b59} Sawicki M., Stevenson
M., Barrientos L. F., Gladman B., Mall\'{e}n-Ornelas G., van den Bergh S.,
2005, ApJ, 627, 621
\bibitem[\protect\citeauthoryear{Schlegeletal}{1998}]{b5} Schlegel D. J., 
Finkbeiner D. P., David, M., 1998, ApJ, 500, 525
\bibitem[\protect\citeauthoryear{Scoccimarroetal}{2001}]{b64} Scoccimarro R., 
Sheth R. K., Hui L., Jain B., 2001, ApJ, 546, 20
\bibitem[\protect\citeauthoryear{Scrantonetal}{2002}]{b64} Scranton R. et al., 
2002, ApJ, 579, 48
\bibitem[\protect\citeauthoryear{Seljak}{2000}]{b5} Seljak U., 2000, MNRAS, 
318, 203
\bibitem[\protect\citeauthoryear{Simpsonetal}{2006}]{b62} Simpson C. et al.,
2006, MNRAS, 373, L21
\bibitem[\protect\citeauthoryear{Smailetal}{2002}]{b64} Smail I., Owen F. N.,
Morrison G. E., Keel W. C., Ivison R. J., Ledlow M. J., 2002, ApJ, 581, 844
\bibitem[\protect\citeauthoryear{Smithetal}{2002}]{b64} Smith G. P. et al., 
2002, MNRAS, 330, 1
\bibitem[\protect\citeauthoryear{Smithetal}{2003}]{b64} Smith R. E. et al., 
2003, MNRAS, 341, 1311
\bibitem[\protect\citeauthoryear{Springel}{2005}]{b5} Springel, 
V. et~al. 2005, Nature, 435, 629
\bibitem[\protect\citeauthoryear{Suraceetal}{2005}]{b64} Surace J., Shupe D. 
L., Fang F., Lonsdale C. J., Gonzalez-Solares E., 2005, SSC Web site Release: 
The SWIRE Data Release 2 (Pasadena, CA: CalTech)
\bibitem[\protect\citeauthoryear{Tinkeretal}{2010}]{b67} Tinker J. L., 
Robertson B. E., Kravtsov A. V., Klypin A., Warren M. S., Yepes G., 
Gottl\"{o}ber S., 2010, ApJ, 724, 878
\bibitem[\protect\citeauthoryear{Tinkeretal}{2010}]{b67} Tinker J. L., 
Weinberg D. H., Zheng Z., Zehavi I., 2005, ApJ, 631, 41
\bibitem[\protect\citeauthoryear{Trichasetal}{2010}]{b5} Trichas M. et al., 
2010, MNRAS, 405, 2243
\bibitem[\protect\citeauthoryear{Tonryetal}{2012}]{b5} Tonry J. L. et al., 
2012, ApJ, 750, 99
\bibitem[\protect\citeauthoryear{vanDokkumetal}{2009}]{b69} van Dokkum P. G.
et al., 2009, PASP, 121, 2
\bibitem[\protect\citeauthoryear{vanDokkumetal}{2010}]{b70} van Dokkum P. G.
et al., 2010, ApJ, 709, 1018
\bibitem[\protect\citeauthoryear{Wakeetal}{2008a}]{b57} Wake D. A. et al., 
2008a, MNRAS, 387, 1045
\bibitem[\protect\citeauthoryear{Wakeetal}{2008b}]{b57} 
Wake, D. A., Croom, S. M., Sadler, E. M., Johnston, H. M.,
2008b, MNRAS, 391, 1674
\bibitem[\protect\citeauthoryear{Wakeetal}{2011}]{b57} Wake D. A. et al., 
2011, ApJ, 728, 46
\bibitem[\protect\citeauthoryear{Whitakeretal}{2011}]{b57} Whitaker K. E. et 
al., 2011, ApJ, 735, 86
\bibitem[\protect\citeauthoryear{WhiteandRees}{1978}]{b57} White S. D.M., 
Rees M. J., 1978, MNRAS, 183, 341
\bibitem[\protect\citeauthoryear{Wright}{2006}]{b57} Wright E. L., 2006, PASP, 
118, 1711
\bibitem[\protect\citeauthoryear{Yagietal}{2002}]{b57} Yagi M., Kashikawa N., 
Sekiguchi M., Doi M., Yasuda N., Shimasaku K., Okamura S., 2002, AJ, 123, 66
\bibitem[\protect\citeauthoryear{Zehavietal}{2002}]{b57} Zehavi I. et al., 
2002, ApJ, 571, 172
\bibitem[\protect\citeauthoryear{Zehavietal}{2005}]{b57} Zehavi I. et al., 
2005, ApJ, 630, 1
\bibitem[\protect\citeauthoryear{Zehavietal}{2011}]{b57} Zehavi I. et al., 
2011, ApJ, 736, 59
\bibitem[\protect\citeauthoryear{Zhengetal}{2005}]{b57} Zheng Z. et al., 2005, 
ApJ, 633, 791
\bibitem[\protect\citeauthoryear{Zhengetal}{2009}]{b57} Zheng Z., Zehavi I., 
Eisenstein D. J., Weinberg D. H., Jing Y. P., 2009, ApJ, 707, 554

\end{thebibliography}
\end{document}